\documentclass[english, aps, pre, reprint, notitlepage, superscriptaddress, longbibliography, twocolumn]{revtex4-2}
\usepackage[T1]{fontenc}
\usepackage[latin9]{inputenc}
\usepackage{amsbsy}
\usepackage{amstext}
\usepackage{amsmath}
\usepackage{amssymb}
\usepackage{graphicx}
\usepackage{esint}
\usepackage{babel}
\usepackage{xcolor}
\usepackage[normalem]{ulem}
\usepackage{array} 
\usepackage[displaymath]{lineno}

\makeatletter
\let\LN@align\align
\let\LN@endalign\endalign
\renewcommand{\align}{\linenomath\LN@align}
\renewcommand{\endalign}{\LN@endalign\endlinenomath}
\let\LN@gather\gather
\let\LN@endgather\endgather
\renewcommand{\gather}{\linenomath\LN@gather}
\renewcommand{\endgather}{\LN@endgather\endlinenomath}
\makeatother

\usepackage[breaklinks=true]{hyperref}
\usepackage{breakcites}
\usepackage{breakurl}

\newcommand{\threshold}{\Theta}

\begin{document}

\title{Infection-induced Cascading Failures -- Impact and Mitigation}

\author{Bo Li}
\email{libo2021@hit.edu.cn}
\affiliation{School of Science, Harbin Institute of Technology (Shenzhen), Shenzhen, 518055, China}
\affiliation{Non-linearity and Complexity Research Group, Aston University, Birmingham,
B4 7ET, United Kingdom}

\author{David Saad}
\email{d.saad@aston.ac.uk}
\affiliation{Non-linearity and Complexity Research Group, Aston University, Birmingham,
B4 7ET, United Kingdom}

\begin{abstract}

\section*{Abstract}

In the context of epidemic spreading, many intricate dynamical patterns can emerge due to the cooperation of different types of pathogens or the interaction between the disease spread and other failure propagation mechanism. To unravel such patterns, simulation frameworks are usually adopted, but they are computationally demanding on big networks and subject to large statistical uncertainty. Here, we study the two-layer spreading processes on unidirectionally dependent networks, where the spreading infection of diseases or malware in one layer can trigger cascading failures in another layer and lead to secondary disasters, e.g., disrupting public services, supply chains, or power distribution. We utilize a dynamic message-passing method to devise efficient algorithms for inferring the system states, which allows one to investigate systematically the nature of complex intertwined spreading processes and evaluate their impact. Based on such dynamic message-passing framework and optimal control, we further develop an effective optimization algorithm for mitigating network failures.
\end{abstract}

\maketitle

\section*{Introduction}
Epidemic outbreaks do not only possess a direct threat to public health but also, indirectly, impact other sectors~\cite{Pak2020, Chaturvedi2021, Cochran2020}. For instance, when many infected individuals have to rest, be hospitalized or quarantined in order to slow down the epidemic spread, this could severely disrupt public services, causing disutility even to those who are not infected. 
For instance, the highly interdependent supply chains can be easily disrupted due to epidemic outbreaks~\cite{Xu2020, Aday2020}. 
Similar concerns apply to cyber security. The spread of malware is not merely detrimental to computer networks, but can also cause failures to power grids or urban transportation networks which rely on modern communication systems~\cite{Amini2019, Liu2020}. What is even worse is that the failures of certain components of technological networks can by themselves  trigger a cascade of secondary failures, which can eventually lead to large-scale outages~\cite{Guo2017}.
Therefore, it is vital to understand the nature of epidemic (or malware) spreading and failure propagation on interacting networks, based on which further mitigation and control measures can be devised.

A number of previous papers address the scenario of  interacting spreading processes. In the context of epidemic spreading, two types of pathogens can cooperate or compete with each other, creating many intricate patterns of disease propagation~\cite{Carlos1996, Carlos1999, Karrer2011, Cai2015, Wang2019, HanlinSun2019, Liu2019}. 
For interacting technological networks (e.g., communication and power networks), the failure of components in one network layer will not only affect neighboring parts within the same network, but will also influence the second network layer through the cross-layer connections.
Macroscopic analyses based on simplified models show that such a spreading mechanism can easily result in a catastrophic breakdown of the whole system~\cite{Buldyrev2010, Bashan2013, Valdez2020}.

Most existing research in the area of multi-layer spreading processes employs macroscopic approaches, such as the degree-distribution-based mean-field methods and asymptotic percolation analysis, in order to obtain the global picture of the models' behavior~\cite{Satorras2015}. Such methods typically do not consider specific network instances and lack the ability to treat the interplay between the spreading dynamics and the fine-grained network topology~\cite{Satorras2015}. For stochastic spreading processes with specific system conditions (e.g., topology initial conditions and individual node properties), it is common to apply extensive Monte Carlo (MC) simulations to observe the evolution of the spread, based on which important policy decisions are made~\cite{Adam2020}. However, such simulations are computationally demanding on big networks and can be subject to large statistical uncertainty; as a result, they are difficult to be used for downstream analysis or optimization tasks. 
Therefore, researchers have been pursuing tractable and accurate theoretical methods to tackle the complex stochastic dynamics on networks~\cite{Satorras2015, Wang2017}. 

Among the various developed theoretical approaches used, dynamic message-passing (DMP) is based on ideas from statistical physics offering a desirable algorithmic framework for approximate inference while it remains computationally efficient~\cite{Karrer2010, Lokhov2014, Lokhov2015}. The DMP method has been shown to be more accurate than the widely adopted individual-based mean-field method, especially in sparse networks~\cite{Koher2019, LiPRE2021}.
Moreover, the DMP approach yields a set of closed-form equations, which is very convenient for additional parameter estimation and optimization tasks~\cite{LokhovNIPS2016, Lokhov2017, HanlinSun2019}.

In this work, we study a scenario where the epidemic or malware spreading on one network can trigger cascading failures on another. This is relevant in the cases where epidemic outbreaks cause disruption in public services or economic activities. Similarly, it can also be applied to study the effect of malware spread on computer networks causing the breakdown of other technological networks such as the power grid. The latter phenomenon is gaining more and more attention due to the increasing interactions among various engineering networks~\cite{Liu2020}.
We explore the dynamics and consequences of such infection-induced cascading failures across two-layer networks using the DMP method. Our results reveal that even relatively low infection rates can induce large-scale cascading failures, leading to widespread network disruptions. We characterized these phenomena through the derivation and analysis of DMP equations, achieving a comprehensive understanding by linking the process to combined bond and bootstrap percolation models analytically. Leveraging the analytical tractability of the DMP model, we also developed optimization algorithms that effectively mitigate these network failures. By adjusting control parameters based on the back-propagation of final state impacts, these algorithms help minimize the size of system failure.

\section*{Methods }
\subsection*{Model and Framework}\label{sec:model_framework}

\subsubsection*{The Model}\label{subsec:model}
\begin{figure}
\includegraphics[scale=0.4]{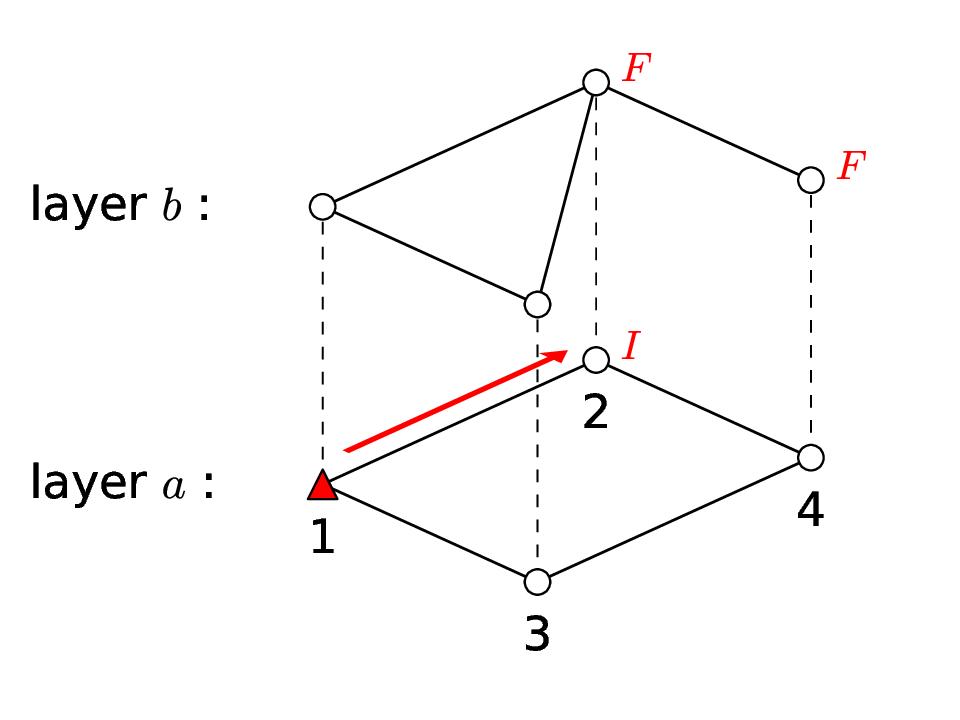}
\caption{An example of the two-layer spreading process considered in this work. 
A node is in state $I$ if it is infected in layer $a$, and a node is in state $F$ if it fails in layer $b$.
In this example, node $2$ is infected by node $1$ in layer $a$, therefore it turns into state $F$ in layer $b$. If $b_{24} \geq \threshold_4$, then node $4$ will also fail as it loses the support from node $2$, even though node $4$ itself has not been infected.
\label{fig:two_layer_net_demo}}
\end{figure}

To study the impact of infection spread of diseases or malware and their secondary effects, we consider multiplex networks comprising two layers~\cite{Boccaletti2014}, which are denoted as layers $a$ and $b$, and are represented by two graphs $G_a(V_a, E_a)$ and $G_b(V_b, E_b)$.
For convenience, we assume that the nodes in both layers correspond to the same set of individuals, denoted as $V = V_a = V_b$. This can be extended to more general settings. Denote $\partial_i^a$ and $\partial_i^b$ as the sets of nodes adjacent to node $i$ in layers $a$ and $b$, respectively. We also define $\partial_i = \partial_i^a \cup \partial_i^b$. See Fig.~\ref{fig:two_layer_net_demo} for an example of the network model under consideration. 

Each node has states on both layers $a$ and $b$.
In layer $a$, each node assumes one of four states, susceptible ($S$), infected ($I$), recovered ($R$), and protected ($P$) at any particular time step. The infection spreading process occurs in layer $a$ only, which is dictated by the stochastic discrete-time SIR model~\cite{Satorras2015} augmented with a protection mechanism, which we term the SIRP model. 
Stochastic models are commonly employed for modeling the spreads of epidemics or malware~\cite{Balcan2010, Adam2020, Garetto2003}.
The stochastic SIR model is commonly used for representing the spread of infections, wherein a susceptible individual (in state $S$) may become infected through contact with infected neighbors, and an infected individual (in state $I$) can recover, transitioning to the recovered state ($R$) after a certain period.
The process we consider is based on the SIR model but includes one more state, $P$,  in layer $a$; it admits the following state-transition rule
\begin{align}
S(i)+I(j) & \xrightarrow{\beta_{ji}} I(i)+I(j),\nonumber \\
I(i) & \xrightarrow{\mu_{i}} R(i),\label{eq:SIRP_def} \\
S(i) & \xrightarrow{\gamma_{i}(t)} P(i),\nonumber
\end{align}
where $\beta_{ji}$ is the probability that node $j$ being in the infected state transmits the infection to its susceptible neighboring node $i$ at a certain time step. At each time step, an existing infected node $i$ recovers with probability $\mu_{i}$; the recovery process is assumed to occur after possible transmission activities. At time $t$, an existing susceptible node $i$ turns into state $P$ if it receives protection at time $t-1$, which occurs with probability $\gamma_i(t-1)$. The protection can be achieved by vaccination in the epidemic setting or special protection measures in the malware spread setting, which is usually subject to certain budget constraints. The protection probabilities $\{ \gamma_i(t) \}$ will be the major control variables for mitigating the outbreaks.
Note that when no protection is provided, i.e., all $\{ \gamma_i(t) \}$ are zero, the SIRP model reduces to the traditional SIR model.
At initial time $t=0$, we assume that node $i$ has a probability $P_S^i(0)$ to be in state $S$, and probability $P_I^i(0) = 1 - P_S^i(0)$ to be in state $I$.

In layer $b$, each node $i$ can either be in the normal state ($N$) or the failed state ($F$), indicated by a binary state variable $x_i$ where $x_i = 1 \; (0)$ denotes the `fail' (`normal') state at a particular time step. A node $i$ in layer $b$ fails if (i) it has been infected, i.e., node $i$ is in state $I$ or $R$ in layer $a$; (ii) there exists certain neighboring failed nodes such that $\sum_{j \in \partial_i^b} x_j b_{ji} \geq \threshold_i$, where $\threshold_i$ is a threshold and the influence parameter $b_{ji}$ measures the importance of the failure of node $j$ on node $i$.
The latter case indicates that node $i$ can fail due to the failures of its neighbors which it relies on, even though node $i$ itself is not infected.
In summary, the failure propagation process in layer $b$ can be expressed as
\begin{align}
x_i = \begin{cases} & \!\!\!\!\! 1, \;\; \text{either (i) node $i$ in state $I$ or $R$ in layer $a$,} \\ 
& \qquad \; \text{or (ii) } \sum_{j \in \partial_i^b} x_j b_{ji} \geq \threshold_i \text{ in layer $b$};  \\
& \!\!\!\!\! 0, \;\; \text{otherwise.}
\end{cases}
\end{align}
The whole process is simulated for $T$ time steps.
As we are interested in the time scale of infection spread which is usually very fast, we do not consider any repair rule in layer $b$. Therefore, a failed node cannot return to normality within the time window under consideration. 

Such a failure propagation mechanism is equivalent to the linear threshold model (LTM), which is commonly used in studying social contagion and other cascade processes~\cite{Watts2002, Kempe2003, Satorras2015}. 
The LTM model also offers a straightforward yet effective framework for understanding cascading failures in various systems, as it effectively encapsulates the pivotal dynamics where a component can become dysfunctional if a significant number of its dependent components fail~\cite{Watts2002, Valdez2020}.
Other popular models for cascading failures incorporate more details of the system functionalities~\cite{Motter2002, Carreras2002, Crucitti2004}; these models require theoretical analyses specific to each case, which fall outside the scope of the current study.

Fig.~\ref{fig:two_layer_net_demo} illustrates the infection-induced cascades of our model in a simple network of 4 nodes. Node $1$ is the initial infected node (or the seed) in layer $a$, which transmits the infection to node $2$ at a certain time step. Now that node $2$ is in the infected state in layer $a$, it also fails to function in layer $b$. If $b_{24} \geq \threshold_4$, then node $4$ will also fail as it loses the support from node $2$, even though node $4$ itself has not been infected. Such additional cascade propagation needs extra care when infections spread out.
Similar interacting SIR (without a protection mechanism) and LTM processes have also been considered in the social contagion setting~\cite{Su2018}.

We reiterate that the infection-spreading process (described by the SIRP model) occurs in layer $a$ only and not the entire network, while the cascade process (described by the LTM model) occurs in layer $b$. 
Typically, a holistic treatment of the combined two-layer processes is needed to understand their impact and develop mitigation strategies.
We also remark that our model differs from the traditional settings of interdependent networks, which typically includes reciprocal dependency.

\subsubsection*{The DMP Framework}\label{subsec:framework}
We aim to use the DMP approach to investigate the two-layer spreading processes described above.
The DMP equations of the usual SIR and the LTM model have been derived, based on the microscopic dynamic belief propagation equations~\cite{Lokhov2015, Altarelli2013}.
As in generic belief propagation methods~\cite{Mezardbook2009}, the DMP method is exact for tree graphs, while it can constitute a good approximation for loopy graphs, particularly when short loops, such as those spanning 3 or 4 nodes, are scarce.
The two-layer spreading processes combining the SIR and LTM model appear more involved, where approximations relying on uncorrelated multiplex networks were used~\cite{Su2018}. Such approximations become less adequate when the two network layers are correlated, e.g., both layers share the same network topology.

\subsubsection*{Dynamic Belief Propagation}
To devise more accurate DMP equations for general network models and accommodate the protection mechanism for mitigation, we start from the principled dynamic belief propagation equations of the two-layer processes. One important characteristic of our model is that state transition is unidirectional, which can only take the direction $S \to I \to R$ or $S \to P$ in layer $a$, and $N \to F$ in layer $b$.
Note that layer $b$ does not influence layer $a$.
As a result, our model admits a reduced representation of the system's dynamical trajectories that subsequently facilitates a drastic simplification of the derivation of the DMP equations, which are exact on tree networks~\cite{Lokhov2015}.
Nevertheless, we emphasize that the exactness of the DMP formalism for tree networks is conditioned on the unidirectional nature of the model, which no longer holds if layer $b$ also influences layer $a$.
Introducing reciprocal interactions between both model layers requires additional theoretical tools, which are interesting by themselves but are beyond the scope of the current study.

Following previous works~\cite{Altarelli2013, Lokhov2015}, we parametrize the dynamical trajectory of each node by its state transition times. In layer $a$, we denote $\tau_i^a, \omega_i^a$ and $\varepsilon_i^a$ as the first time at which node $i$ turns into state $I$, $R$ and $P$, respectively. In layer $b$, we denote $\tau_i^b$ as the first time at which node $i$ turns into state $F$.
The quantity of interest is the probability of the trajectory of node $i$ considered in the entire graph comprising layers $a$ and $b$ but having a cavity where node $j$ is absent, denoted as $m^{i \to j}(\tau_{i}^{a},\omega_{i}^{a},\varepsilon_{i}^{a},\tau_{i}^{b})$.
Throughout the manuscript, we will refer to probabilities defined within a cavity graph as cavity probabilities.
It is computed by the following dynamic belief propagation equations
\begin{align}
& \qquad m^{i \to j}(\tau_{i}^{a},\omega_{i}^{a},\varepsilon_{i}^{a},\tau_{i}^{b}) \nonumber \\
= & \!\!\! \sum_{\{\tau_{k}^{a},\omega_{k}^{a},\varepsilon_{k}^{a},\tau_{k}^{b}\}_{k\in\partial_{i}}} \!\!\!\!\!\!\!\!\! W_{\text{SIRP}}^{i}(\tau_{i}^{a},\omega_{i}^{a},\varepsilon_{i}^{a}||\{\tau_{k}^{a},\omega_{k}^{a},\varepsilon_{k}^{a}\}_{k\in\partial_{i}^{a}}) \nonumber \\
& \times W_{\text{LTM}}^{i}(\tau_{i}^{b}||\tau_{i}^{a},\varepsilon_{i}^{a},\{\tau_{k}^{b}\}_{k\in\partial_{i}^{b}}) \nonumber \\
& \times \prod_{k\in\partial_{i} 
\backslash j }m^{k\to i}( \tau_{k}^{a},\omega_{k}^{a},\varepsilon_{k}^{a},\tau_{k}^{b} ), \label{eq:DBP_m_itoj}
\end{align}
where $W_{\text{SIRP}}^{i}(\cdot)$ and $W_{\text{LTM}}^{i}(\cdot)$ are the transition kernels dictated by the dynamical rules of the SIRP and LTM model, respectively (for details see Supplementary Note 1). The marginal probability of the trajectory of node $i$, denoted as $m^{i}(\tau_{i}^{a},\omega_{i}^{a},\varepsilon_{i}^{a},\tau_{i}^{b})$, can be computed in a similar way as Eq.~(\ref{eq:DBP_m_itoj}), by replacing the product $\prod_{k\in\partial_{i} 
\backslash j}$ in the last line of Eq.~(\ref{eq:DBP_m_itoj}) by $\prod_{k\in\partial_{i}}$.
That is, the marginal probability $m^{i}(\cdot)$ is calculated using the entire graph, in contrast to the cavity probability $m^{i \to j}(\cdot)$ which is determined with a cavity graph where node $j$ is absent.

The  probability of node $i$ in a certain state can be computed by summing the trajectory-level probability, which will be described in the next section.

\subsection*{Full Node-level DMP Equations}\label{sec:node_level_DMP}
Consider the cavity probability of node $i$ being in state $S$ in layer $a$ at time $t$ (assuming node $j$ is absent - the cavity), it is obtained by tracing over the corresponding probabilities of trajectories $m^{i \to j}( \cdot )$ in the cavity graph (assuming node $j$ is removed)
\begin{align}
P_{S}^{i\to j}(t) = \sum_{\tau_{i}^{a}, \omega_{i}^{a}, \varepsilon_{i}^{a}, \tau_{i}^{b} } & \mathbb{I}(t<\tau_{i}^{a}<\omega_{i}^{a}) \mathbb{I}( t <\varepsilon_{i}^{a} ) \nonumber \\
& \times m^{i\to j} (\tau_{i}^{a},\omega_{i}^{a},\varepsilon_{i}^{a}, \tau_{i}^{b}), \label{eq:PS_cav}
\end{align}
where $\mathbb{I}(\cdot)$ is the indicator function enforcing the order of state transitions. Similarly, we denote the cavity probability of node $i$ in state $F$ in layer $b$ (in the absence of node $j$) as $P_F^{i\to j}(t)$; it is obtained by
\begin{align}
P_F^{i \to j}(t) = \sum_{\tau_{i}^{a}, \omega_{i}^{a}, \varepsilon_{i}^{a}, \tau_{i}^{b} }  \mathbb{I}(\tau_{i}^{b} \leq t) m^{i\to j} (\tau_{i}^{a},\omega_{i}^{a},\varepsilon_{i}^{a}, \tau_{i}^{b}).  \label{eq:q_cav}
\end{align}
The marginal probabilities $P_S^i(t)$ and $P_F^i(t)$ can be computed in a similar manner, by replacing $m^{i \to j}(\cdot)$ in Eq.~(\ref{eq:PS_cav}) and Eq.~(\ref{eq:q_cav}) with $m^i(\cdot)$.

\subsubsection*{DMP Equations in Layer $a$}\label{sec:DMP_layer_a}
We note that infection spread in layer $a$ is not influenced by cascades in
layer $b$, while the failure time in layer $b$ depends on the infection time and the protection time of the corresponding node in layer $a$. Hence, we can decompose the message $m^{i \to j}(\cdot)$ to the respective components as
\begin{align}
m^{i \to j}(\tau_{i}^{a},\omega_{i}^{a}, \varepsilon_{i}^{a},\tau_{i}^{b}) = \;\; &  m_{a}^{i \to j}(\tau_{i}^{a}, \omega_{i}^{a}, \varepsilon_{i}^{a}) \nonumber \\
& \times m_{b}^{i \to j}(\tau_{i}^{b} \mid \tau_{i}^{a}, \varepsilon_{i}^{a}).
\end{align}
where $m_a^{i \to j}(\cdot)$ and $m_b^{i \to j}(\cdot)$ denote the trajectory-level probabilities of the processes in layer $a$ and $b$, respectively.
Note that the messages $\{ m^{i \to j}(\cdot) \}$ live in the entire network comprising layers $a$ and $b$, which implies that $\{ m_a^{i \to j}(\cdot), m_b^{i \to j}(\cdot) \}$ are also defined on the entire network.

Summing $m_{a}^{i \to j}(\cdot)$ over $\tau_{i}^{a},\omega_{i}^{a}, \varepsilon_{i}^{a}$ up to a certain time yields the normal DMP equations of node-level probabilities for the infection spread in layer $a$ (see details in Supplementary Note 1). They admit the following expressions for $t > 0$
\begin{align}
& P_{S}^{i\to j}(t)=P_{S}^{i}(0)\prod_{t'=0}^{t-1}\big[1-\gamma_{i}(t')\big]\prod_{k\in\partial_i^a \backslash j}\theta^{k\to i}(t), \label{eq:PS_ki_t_DMP_SIRP} 	\\
& \theta^{k\to i}(t) = \theta^{k\to i}(t-1)-\beta_{ki}\phi^{k\to i}(t-1), \label{eq:theta_ki_t_iter_SIRP}  \\
& \phi^{k\to i}(t) = \big(1-\beta_{ki}\big)\big(1-\mu_{k}\big)\phi^{k\to i}(t-1) \label{eq:phi_ki_t_iter_SIRP} \nonumber \\
 & \qquad - \bigg\{ P_{S}^{k\to i}(t)-P_{S}^{k\to i}(t-1)\big[1-\gamma_{k}(t-1)\big]\bigg\}, 
\end{align}
where $\theta^{k \to i}(t)$ is the cavity probability that node $k$ has not transmitted the infection signal to node $i$ up to time $t$, and $\phi^{k \to i}(t)$ is
the cavity probability that $k$ is in state $I$ but has not transmitted the infection
signal to node $i$ up to time $t$.
Note that the messages $\{ P_{S}^{k\to i}(t), \theta^{k\to i}(t), \phi^{k\to i}(t) \}$ are only needed for edges belonging to layer $a$ where the SIRP model is defined.

At time $t=0$, as we consider that each node $i$ is either in state $S$ with probability $P_S^i(0)$ or in state $I$ with probability $1 - P_S^i(0)$, we have the following initial conditions for the messages
\begin{align}
P_{S}^{i\to j}(0) & =P_{S}^{i}(0), 	\nonumber\\
\phi^{i\to j}(0) & =1-P_{S}^{i}(0),\nonumber\\
\theta^{i\to j}(0) & =1.\label{eq:DMPinitial}
\end{align}

Upon iterating the above messages~(\ref{eq:PS_ki_t_DMP_SIRP})-(\ref{eq:phi_ki_t_iter_SIRP})  starting from the initial conditions~(\ref{eq:DMPinitial}), the node-level marginal probabilities can be computed as
\begin{align}
& P_{S}^{i}(t)=P_{S}^{i}(0)\prod_{t'=0}^{t-1}\big[1-\gamma_{i}(t')\big]\prod_{k\in\partial_i^a}\theta^{k\to i}(t),\label{eq:PS_i_t_DMP_SIRP}  \\
& P_{R}^{i}(t)= P_{R}^{i}(t-1)+\mu_{i}P_{I}^{i}(t-1),\label{eq:PR_DMP_iter_SIRP} \\
& P_{P}^{i}(t)= P_{P}^{i}(t-1)+\gamma_{i}(t-1)P_{S}^{i}(t-1),\label{eq:PP_DMP_iter_SIRP}\\
& P_{I}^{i}(t)= 1-P_{S}^{i}(t)-P_{R}^{i}(t)-P_{P}^{i}(t).\label{eq:PI_DMP_iter_SIRP}
\end{align}

The above DMP equations~(\ref{eq:PS_i_t_DMP_SIRP})-(\ref{eq:PI_DMP_iter_SIRP}) bear similarity to those of SIR model~\cite{Lokhov2014}, except for the protection mechanism with control parameters $\{ \gamma_i(t) \}$.
The computational complexity for obtaining the messages for the SIRP process in layer $a$ over a total time $T$ is $O(|E_a|T)$, where $|E_a|$ denotes the number of edges in layer $a$.

\subsubsection*{DMP Equations in Layer $b$}\label{sec:DMP_layer_b}
As for the cascade process in layer $b$, whether node $i$ will turn into state $F$ (fail) also depends on the state in layer $a$, making it more challenging to derive the corresponding DMP equations. The key to obtaining node-level DMP equations for $P_F^{i \to j}(t)$ in Eq.~(\ref{eq:q_cav}) (and the corresponding marginal probability $P_F^i(t)$) is to introduce several intermediate quantities to facilitate the calculation; the details are outlined in Supplementary Note 1.

To summarize, the node-level failure probability $P_F^i(t)$ can be decomposed as
\begin{align}
P_F^i(t) & = P_I^i(t) + P_R^i(t) + P_{SF}^i(t) + P_{PF}^i(t),
\end{align}
where $P_{SF}^i(t)$ and $P_{PF}^i(t)$ are the probabilities that node $i$ is in state $F$ in layer $b$, while it is in state $S$ or state $P$ in layer $a$, respectively. For these two cases, the failure of node $i$ is triggered by the failure propagation of its neighbors from layer $b$. A similar relation holds for the cavity probability $P_F^{i \to j}(t)$.

The probability $P_{SF}^i(t)$ admits the following iteration
\begin{align}
& P_{SF}^{i}(t) = P_{S}^{i}(0)\prod_{t'=0}^{t-1}\big[1-\gamma_{i}(t')\big] \prod_{k\in\partial_{i}^{a}\backslash\partial_{i}^{a}\cap\partial_{i}^{b}}\theta^{k\to i}(t) \nonumber \\
& \times \sum_{\{x_{k}\}_{k\in\partial_{i}^{b}}}\mathbb{I}\bigg(\sum_{k\in\partial_{i}^{b}} x_{k} b_{ki}  \geq\Theta_{i}\bigg)\nonumber \\
& \times\prod_{ \substack{ k\in\partial_{i}^{b}\backslash\partial_{i}^{a}\cap\partial_{i}^{b}, \\  x_{k}=1} }P_F^{k\to i}(t-1) \prod_{ \substack{ k\in\partial_{i}^{b}\backslash\partial_{i}^{a}\cap\partial_{i}^{b}, \\ x_{k}=0} }\big[1-P_F^{k\to i}(t-1)\big]\nonumber \\
& \times\prod_{ \substack{  k\in\partial_{i}^{a}\cap\partial_{i}^{b}, \\ x_{k}=1 } } \chi^{k\to i}(t) \prod_{ \substack{ k\in\partial_{i}^{a}\cap\partial_{i}^{b}, \\ x_{k}=0}  } \big[ \theta^{k\to i}(t) - \chi^{k\to i}(t) \big], \label{eq:P_SF_i}
\end{align}
where $\chi^{k\to i}(t)$ is the cavity probability that node $k$ is in state $F$ at time $t-1$, and it has not sent the infection signal to node $i$ up to time $t$.

The cavity probability $\chi^{k\to i}(t)$ can be decomposed into
\begin{align}
\chi^{k\to i}(t) = \psi^{k\to i}(t) + P_{SF}^{k\to i}(t-1) + P_{PF}^{k\to i}(t-1),
\end{align}
where $\psi^{k\to i}(t)$ is the cavity probability that node $k$ is in state $I$ or $R$ at time $t-1$, but has not transmitted the infection signal to node $i$ up to time $t$. The cavity probability $\psi^{k\to i}(t)$ can be computed as
\begin{align}
& \psi^{k\to i}(t) = \psi^{k\to i}(t-1) - \beta_{ki}\phi^{k\to i}(t-1) \nonumber \\
& \qquad + \big[1-\gamma_{k}(t-2)\big]P_{S}^{k\to i}(t-2)-P_{S}^{k\to i}(t-1).
\end{align}

Similarly, the probability $P_{PF}^i(t)$ admits the following iteration
\begin{align}
& P_{PF}^{i}(t)= P_{S}^{i}(0) \sum_{\varepsilon=1}^{t}\gamma_{i}(\varepsilon-1)\prod_{t'=0}^{\varepsilon-2}\big[1-\gamma_{i}(t')\big] \label{eq:P_PF_i}  \\
& \times \prod_{k\in\partial_{i}^{a}\backslash\partial_{i}^{a}\cap\partial_{i}^{b}}\theta^{k\to i}(\varepsilon-1)  \sum_{\{x_{k}\}_{k\in\partial_{i}^{b}}} \mathbb{I}\bigg(\sum_{k\in\partial_{i}^{b}} x_{k} b_{ki} \geq\Theta_{i}\bigg)\nonumber \\
& \times \prod_{ \substack{ k\in\partial_{i}^{b}\backslash\partial_{i}^{a}\cap\partial_{i}^{b}, \\ x_{k}=1 } } P_F^{k\to i}(t-1) \prod_{  \substack{ k\in\partial_{i}^{b}\backslash\partial_{i}^{a}\cap\partial_{i}^{b}, \\ x_{k}=0 } }\big[1-P_F^{k\to i}(t-1)\big]\nonumber \\
& \times \prod_{ \substack{ k\in\partial_{i}^{a}\cap\partial_{i}^{b}, \\ x_{k}=1} } \tilde{\chi}^{k\to i}(t,\varepsilon) \prod_{ \substack{ k\in\partial_{i}^{a}\cap\partial_{i}^{b}, \\ x_{k}=0 } }\big[\theta^{k\to i}(\varepsilon-1)-\tilde{\chi}^{k\to i}(t,\varepsilon)\big], \nonumber
\end{align}
where the dummy variable $\varepsilon$ indicates the time at which node $i$ receives the protection signal. 

In Eq.~(\ref{eq:P_PF_i}), $\tilde{\chi}^{k\to i}(t,\varepsilon)$ is the cavity probability that node $k$ is in state $F$ at time $t-1$, but has not transmitted the infection signal to node $i$ up to time $\varepsilon$. It can be decomposed into
\begin{align}
\tilde{\chi}^{k\to i}(t, \varepsilon) = \tilde{\psi}^{k\to i}(t, \varepsilon) + P_{SF}^{k\to i}(t-1) + P_{PF}^{k\to i}(t-1),
\end{align}
where $\tilde{\psi}^{k\to i}(t, \varepsilon)$ is the cavity probability that node $k$ is in state $I$ or $R$ at time $t-1$, but has not transmitted the infection signal to node $i$ up to time $\varepsilon - 1$. The cavity probability $\tilde{\psi}^{k\to i}(t)$ can be computed as
\begin{align}
\tilde{\psi}^{k\to i}(t, \varepsilon) = \; & \psi^{k\to i}(\varepsilon - 1) + P_{I}^{k\to i}(t-1) + P_{R}^{k\to i}(t-1) \nonumber \\
&  - \left[ P_{I}^{k\to i}(\varepsilon - 2) + P_{R}^{k\to i}(\varepsilon - 2) \right].
\end{align}

Note that the cavity probabilities $P_{SF}^{i \to j}(t)$ and $P_{PF}^{i \to j}(t)$ are computed using the similar formula as in Eq.~(\ref{eq:P_SF_i}) and Eq.~(\ref{eq:P_PF_i}), but in the cavity graph where node $j$ is removed. This closes the loop for the DMP equations in layer $b$.
We also observe in the above equations that the node-level messages for the SIRP process only enter into the DMP equations for the LTM process through the overlapping neighbors $\partial_i^a \cap \partial_i^b$.

The initial conditions for the corresponding messages are given by
\begin{align}
& P_F^{k}(0) = P_F^{k \to i}(0) = P_I^k(0), \\
& P_{SF}^{k}(0)=P_{SF}^{k\to i}(0)=0, \\
& P_{PF}^{k}(0)=P_{PF}^{k\to i}(0)=0, \\
& \psi^{k\to i}(1) = \chi^{k\to i}(1) = (1 - \beta_{ki}) P_{I}^{k}(0), \\
& \tilde{\psi}^{k\to i}(1,1) = \tilde{\chi}^{k\to i}(1,1) = P_{I}^{k}(0).
\end{align}
For $t \geq 2, \varepsilon = 1$, we have
\begin{align}
\tilde{\psi}^{k\to i}(t,\varepsilon=1) = \; & P_{I}^{k\to i}(t-1)+P_{R}^{k\to i}(t-1), \\
\tilde{\chi}^{k\to i}(t,\varepsilon=1) = \; & P_{I}^{k\to i}(t-1)+P_{R}^{k\to i}(t-1) \nonumber \\ 
& + P_{SF}^{k\to i}(t-1)+P_{PF}^{k\to i}(t-1).
\end{align}

We remark that for a total time $T$, the computational complexity for obtaining the messages of the cascade process in layer $b$ is $O(|E_b| T^2)$ where $|E_b|$ denotes the number of edges in layer $b$, unlike the $O(|E_a| T)$ complexity for the SIRP process in layer $a$. This is due to the dependency of layer $b$ on layer $a$, as well as the protection mechanism in layer $a$. 
The summation of the dummy state $\{ x_k \}_{k \in \partial_i^b}$ in Eq.~(\ref{eq:P_SF_i}) and Eq.~(\ref{eq:P_PF_i}) also implies a high computational demand of networks with high-degree nodes. One way to alleviate this complexity is to use the dynamic programming techniques introduced in by Torrisi et al.~\cite{Torrisi2021}.

These DMP equations are exact if both layers are tree networks, while they are approximate solutions when there are loops in the underlying networks.

\subsection*{Simplification under Small Inter-layer Overlap}\label{sec:simplify_DMP}
If there are no overlaps between the neighbors of node $i$ in layer $a$ and those in layer $b$, i.e., $\partial_{i}^{a}\cap\partial_{i}^{b} = \varnothing$, the messages $\chi^{k \to i}, \psi^{k \to i}, \tilde{\chi}^{k \to i}$ and $\tilde{\psi}^{k \to i}$ are not needed, and the node-level probabilities $P_{SF}^i(t)$ and $P_{PF}^i(t)$ can be much simplified as
\begin{align}
& P_{SF}^{i}(t) = P_{S}^{i}(t)\sum_{\{x_{k}\}_{k\in\partial_{i}^{b}}}\mathbb{I}\bigg(\sum_{k\in\partial_{i}^{b}} x_{k} b_{ki} \geq\Theta_{i}\bigg)  \\
& \quad \times\prod_{k\in\partial_{i}^{b},x_{k}=1}P_F^{k\to i}(t-1)\prod_{k\in\partial_{i}^{b},x_{k}=0}\big[1-P_F^{k\to i}(t-1)\big], \nonumber \\
& P_{PF}^{i}(t)= P_{P}^{i}(t)\sum_{\{x_{k}\}_{k\in\partial_{i}^{b}}}\mathbb{I}\bigg(\sum_{k\in\partial_{i}^{b}} x_{k} b_{ki} \geq\Theta_{i}\bigg)  \\
& \quad \times\prod_{k\in\partial_{i}^{b},x_{k}=1}P_F^{k\to i}(t-1)\prod_{k\in\partial_{i}^{b},x_{k}=0}\big[1-P_F^{k\to i}(t-1)\big]. \nonumber
\end{align}
This is also a reasonable approximation if the two layers $a$ and $b$ have little correlation, which has been exploited by previous work~\cite{Su2018}.
We remark that the computational complexity of obtaining messages for the cascade process in layer $b$ using this approximated method is $O(|E_b| T)$.
In this work, we will employ this approximation when we consider the dynamics in the large time limit and devise an optimization algorithm for mitigating the cascading failures, in order to reduce computing time.
In situations where inter-layer overlaps are significant and accuracy is important~\cite{Parshani2010, Cellai2013}, one can always use the complete formulations of the DMP equations as detailed in the ``Full Node-level DMP Equations'' subsection above.

\section*{Results}

\subsection*{Effectiveness of the DMP Method}
\begin{figure}
\hspace{-0.5em}\includegraphics[scale=1.08]{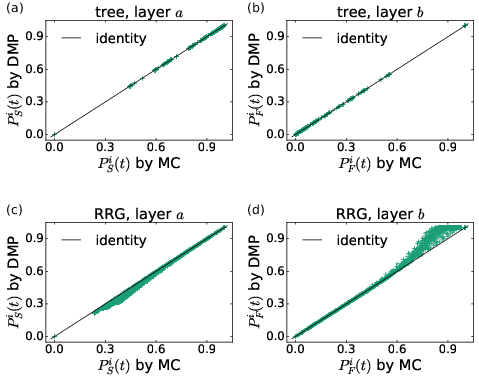}
\caption{Comparison of node-level probabilities. The node-level probabilities $P_S^i(t)$ and $P_F^i(t)$ are obtained by the DMP theory and Monte Carlo (MC) simulation (averaged over $10^5$ realizations). Panels (a) and (b) correspond to a binary tree network of size $N = 63$ for both layers. Panels (c) and (d) correspond to a random regular graph (RRG) of size $N = 100$ and degree $K = 5$ for both layers. The system parameters are $T = 50, \beta_{ji} = 0.2, \mu_i = 0.5, b_{ji} = 1, \Theta_i =  0.6 |\partial_i^b|, \gamma_i(t) = 0$.}
\label{fig:compare_DMP_MC}
\end{figure}

We firstly test the efficacy of the complete DMP equations derived in ``Full Node-level DMP Equations'' subsection in the Methods section, by comparing the node-level probabilities $P_S^i(t)$ and $P_F^i(t)$ to those obtained by Monte Carlo simulations. The DMP theory produces exact marginal probabilities for node activities in tree networks; this is verified in Fig.~\ref{fig:compare_DMP_MC}(a) and (b) where both layers $a$ and $b$ are the same binary tree network of size $N = 63$. For random regular graphs (RRG) where there are many loops, the DMP method also yields reasonably accurate solutions; this is demonstrated in Fig.~\ref{fig:compare_DMP_MC}(c) and (d) where both layers $a$ and $b$ are the same RRG of size $N = 100$ and degree $K = 5$. 
We also validate the effectiveness of the non-overlapping approximation applied to the DMP equations for the process in layer $b$ introduced in the subsection ``Simplification under Small Inter-layer Overlap'' in Methods; the results are shown in Supplementary Note 2.

\subsection*{Impact of Infection-induced Cascades}\label{sec:impact}
The obtained DMP equations of the two-layer spreading processes allow us to examine the impact of the infection-induced cascading failures, on either a specific instance of a multiplex network or an ensemble of networks following a certain degree distribution. In this section, we do not consider the protection of nodes by setting $\gamma_i(t) = 0$, where the process in layer $a$ is essentially a discrete-time SIR model.

\subsubsection*{Impact on a Specific Network}
For the process in layer $a$, we define the outbreak size at time $t$ as the fraction of nodes that have been infected at that time
\begin{align}
\rho_I(t) + \rho_R(t) = \frac{1}{N} \sum_{i \in V_a} P_I^i(t) + \frac{1}{N} \sum_{i \in V_a} P_R^i(t).
\end{align}
For the process in layer $b$, we define the cascade size at time $t$ as the fraction of nodes that have failed at that time
\begin{align}
\rho_F(t) = \frac{1}{N} \sum_{i \in V_b} P_F^i(t).
\end{align}
By definition, we have $\rho_F(t) \geq \rho_I(t) + \rho_R(t)$.

\begin{figure}
\includegraphics[scale=0.37]{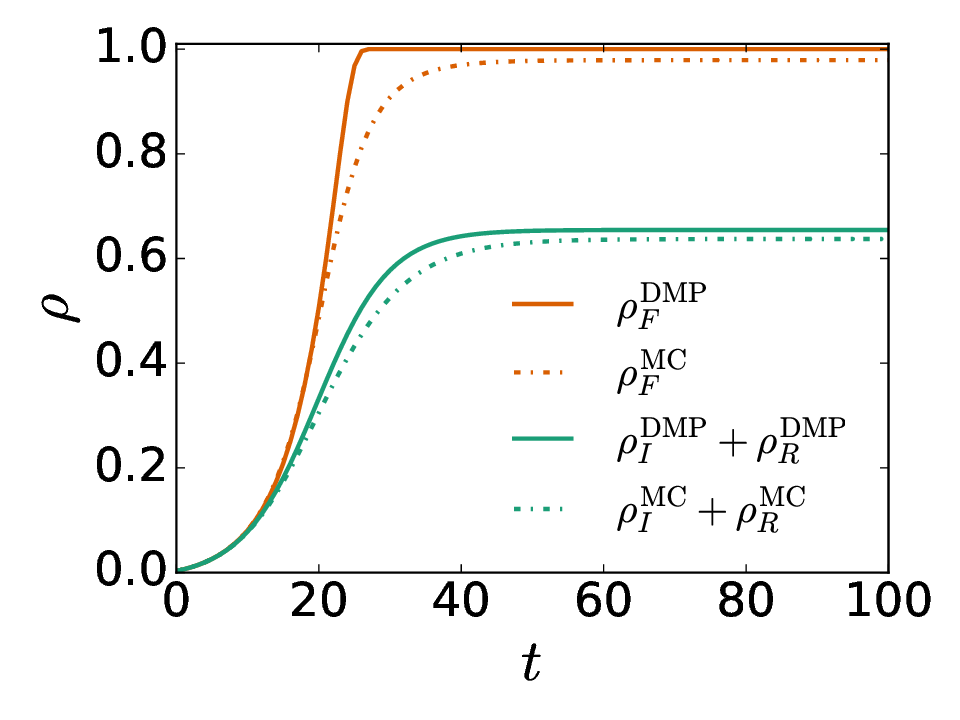}
\caption{Evolution of the sizes of the infection outbreak in layer $a$ and total failures in layer $b$. The size of infection outbreak is measured by $\rho_I + \rho_R$ (green lines), while the size of total failures is measured by $\rho_F$ (orange lines). Both the DMP method (solid line) and MC simulation (dashed-dotted line) are considered. Layer $a$ and layer $b$ have different network topologies, but both are realizations of random regular graphs of size $N=1600$ and degree $K = 5$. At time $t=0$, there are 5 infected nodes. The system parameters are $\beta_{ji} = 0.2, \mu_i = 0.5, b_{ji} = 1, \Theta_i =  0.6 |\partial_i^b|, \gamma_i(t) = 0$.}
\label{fig:outbreak_vs_t_rrg_N1600}
\end{figure}

In Fig.~\ref{fig:outbreak_vs_t_rrg_N1600}, we demonstrate the time evolution of the infection outbreak size and the cascade size in a multiplex network where both layers are random regular graphs with size $N = 1600$. 
It can be observed that $\rho_F$ is much larger than $\rho_I + \rho_R$ asymptotically, which suggests that the failure propagation mechanism in layer $b$ significantly amplifies the impact of the infection outbreaks in layer $a$. In particular, the failure can eventually propagate to the whole network even though less than 70\% of the population gets infected when the spread of the infection saturates. 
Compare to Monte Carlo simulations, the DMP method systematically overestimates the outbreak sizes due to the effect of mutual infection, but it has been shown to offer a significant improvement over the individual-based mean-field method~\cite{Shrestha2015, Koher2019, LiPRE2021}. 

\subsubsection*{Asymptotic Properties}
In the above example, the system converges to a steady state in the large time limit. The DMP approach allows us to systematically investigate the asymptotic behavior of the two-layer spreading processes.

For the process in layer $a$, we define an auxiliary probability
\begin{equation}
p_{ij} := \frac{ \beta_{ij} }{ \beta_{ij} + \mu_i - \beta_{ij} \mu_i}. \label{eq:pij_def}    
\end{equation}
Then the messages in layer $a$ admit the following expressions in the limit $T \to \infty$
\begin{align}
& \phi^{i \to j}(\infty) = 0, \nonumber \\
& \theta^{i \to j}(\infty) = 1 - p_{ij} + p_{ij} P_S^{i \to j}(\infty),  \nonumber \\
& P_S^{i \to j}(\infty) =  P_S^i(0) \prod_{k \in \partial^a_i \backslash j } \theta^{k \to i}(\infty),  \nonumber \\
& P_S^{i}(\infty) =  P_S^i(0) \prod_{k \in \partial^a_i } \theta^{k \to i}(\infty), \label{eq:DMPasymptotic}
\end{align}
Details of the derivation can be found in Supplementary Note 3. The above asymptotic equations~\eqref{eq:DMPasymptotic} suggest a well-known relationship between epidemic spreading and bond percolation~\cite{Grassberger1983, Karrer2010, Satorras2015}.
The bond percolation problem involves a network where the bonds (or edges) between nodes are randomly occupied with a certain probability (denoted as $\lambda$).
The main focus is to understand the formation of a giant cluster comprising connected occupied edges in the network; in large systems, this typically occurs when $\lambda$ is greater than a transition point $\lambda_c$~\cite{MingLi2021}.

As mentioned above, it is well established that the asymptotic properties of many stochastic epidemic spreading models can be mapped to certain bond percolation problems~\cite{Grassberger1983, Newman2002}; we refer interested readers to two recent reviews for more details on the subject~\cite{Satorras2015, MingLi2021}.
In the SIR model studied here (where $\gamma_i(t) = 0$), the quantity $p_{ij}$ defined in Eq.~(\ref{eq:pij_def}) can be interpreted as the probability that an infection transmission on edge $(i,j)$ has been realized in the long run, corresponding to an edge occupation probability in bond percolation.
When the transmission probabilities $\{ \beta_{ij} \}$ are large ($\{ p_{ij} \}$ will also be large), a few initially infected seeds can eventually infect a significant proportion of the population and lead to a pandemic, which corresponds to the formation of a giant cluster in percolation theory. We refer readers to Supplementary Note 3 for more details of the correspondence between our model and bond percolation.
Note that the edge occupation probability $p_{ij}$ in this discrete-time SIR model differs from the continuous-time counterpart~\cite{Grassberger1983, Karrer2010} with an additional term $\beta_{ij} \mu_i$ in the denominator. The term $\beta_{ij} \mu_i$ accounts for the simultaneous events that node $i$ infects node $j$ and recovers within the same time step~\cite{Koher2019}.

For the process in layer $b$, we assume that layers $a$ and $b$ are weakly correlated due to their different topologies and adopt the approximation made in 
the subsection ``Simplification under Small Inter-layer Overlap'' in Methods.
As no protection is applied, we have $P_{PF}^i(t) = 0$. Then the messages in layer $b$ admit the following expression in the limit $T \to \infty$
\begin{align}
& P_F^{i \to j}(\infty)  =  1 - P_S^i(\infty) \label{eq:PF_inf} \\ 
& \quad +  P_S^i(\infty) \sum_{ \{ x_{k} \}_{k\in\partial_{i}^{b}\backslash j} }\mathbb{I}\bigg(\sum_{k\in\partial_{i}^{b}\backslash j} 
 x_{k} b_{ki}\geq\Theta_{i}\bigg) \nonumber  \\
& \quad \quad \times\prod_{k\in\partial_{i}^{b}\backslash j,x_{k}=1}P_F^{k\to i}(\infty)\prod_{k\in\partial_{i}^{b}\backslash j,x_{k}=0}\big[1-P_F^{k\to i}(\infty)\big], \nonumber
\end{align}
where a similar expression holds for $P_F^i(\infty)$ by replacing $\partial_{i}^{b}\backslash j$ with $\partial_{i}^{b}$ in Eq.~(\ref{eq:PF_inf}). The asymptotic equations for layer $b$ suggest a relationship between the LTM model and bootstrap percolation~\cite{Altarelli2013}.

\subsubsection*{Two-layer Percolation in Large Homogeneous Networks}
The large-time behaviors of the two processes correspond to two types of percolation problems. 
To further examine the macroscopic critical behaviors of the two-layer percolation models, it is convenient to consider large-size random regular graphs of degree $K$ (which have a homogeneous network topology), and homogeneous system parameters with $\beta_{ji} = \beta, \mu_i = \mu, b_{ji} = b, \Theta_i = \Theta$.
We further assume that each node $i$ has a vanishingly small probability of being infected at time $t=0$ with $P_I^i(0) = 1 - P_S^i(0) \propto 1 / N$. In the large size limit $N \to \infty$, we have $P_S^i(0) \to 1$. 

Due to the homogeneity of the system, one can assume that all messages and marginal probabilities are identical,
\begin{align}
& \theta^{i \to j}(\infty) = \theta^{\infty}, \\
& P_F^{i \to j}(\infty) = P_F^{\infty}, \\
& P_S^i(\infty) = \rho_S^\infty, \\
& P_F^i(\infty) = \rho_F^\infty. 
\end{align}
It leads to the self-consistent equations in the large size limit ($N \to \infty$),
\begin{align}
& \theta^\infty = 1 - p + p \cdot (\theta^\infty)^{K-1}, \label{eq:theta_inf} \\
& \rho_S^\infty = (\theta^\infty)^{K}, \label{eq:rhoS_inf}  \\
& P_F^{\infty} = 1 - \rho_S^\infty \label{eq:P_F_inf}  \\
& \quad \quad \;\;  + \rho_S^\infty \sum_{n = \lceil \Theta \rceil}^{K -1} \binom{K-1}{n} (P_F^{\infty})^{n} (1 - P_F^{\infty})^{K - 1 - n}, \nonumber \\
& \rho_F^{\infty} = 1 - \rho_S^\infty \label{eq:rhoF_inf}  \\
& \quad \quad \;\;  + \rho_S^\infty \sum_{n = \lceil \Theta \rceil}^{K} \binom{K}{n} (P_F^{\infty})^{n} (1 - P_F^{\infty})^{K - n}, \nonumber
\end{align}
where $p = \frac{\beta}{\beta + \mu - \beta\mu}$ and $\lceil x \rceil$ is the smallest integer greater than or equal to $x$.

We observe that $\theta^{\infty} = 1, \rho_S^\infty = 1, P_F^{\infty} = 0, \rho_F^\infty = 0$ is always a fixed point to Eqs.~(\ref{eq:theta_inf})-(\ref{eq:rhoF_inf}), which corresponds to vanishing outbreak sizes.
When the infection probability $\beta$ is larger than a critical point $\beta_c^a$, this fixed point solution becomes unstable and another fixed point with finite outbreak sizes develops. 

\begin{figure}
\hspace{-1em}\includegraphics[scale=1.1]{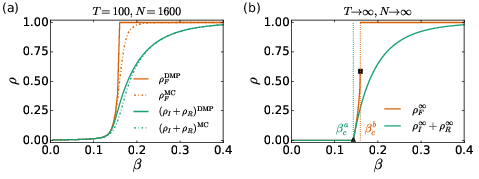}
\caption{Size of infection outbreak and total failures as a function of the infection probability $\beta$. The size of infection outbreak is measured by $\rho_I + \rho_R$ (green lines), while the size of total failures is measured by $\rho_F$ (orange lines). The limits of large system size and large time are considered.  (a) Random regular graphs with $N = 1600, K=5$ are considered. The spreading processes are iterated for $T = 100$ steps, where stationary states are attained. Both the DMP method (solid line) and MC simulation (dashed-dotted line) are considered. (b) Random regular graphs with $K = 5$ in the asymptotic limit $T \to \infty, N \to \infty$ are considered by analyzing the large-time behaviors of the DMP equations. The triangle and the square markers indicate the locations of the two transition points $\beta_c^a$ and $\beta_c^b$, respectively. The system parameters are homogeneous, with $\mu = 0.5, b = 1, \Theta =  3, \gamma_i(t) = 0$.}
\label{fig:outbreak_asymptotic}
\end{figure}

As a concrete example, we consider random regular graphs of degree $K=5$ and fix $\mu = 0.5, b = 1, \Theta = 3$. By solving Eqs.~(\ref{eq:theta_inf})-(\ref{eq:rhoF_inf}) for different $\beta$, we obtain outbreak sizes for both layers $a$ and $b$ under different infection strengths. 
The result is shown in Fig.~\ref{fig:outbreak_asymptotic}, where the asymptotic theory accurately predicts the behavior of a large-size system ($N = 1600$) in the large-time limit.
It is also observed that the outbreak sizes in both layers become non-zero when $\beta$ is larger than a critical point $\beta_c^a = \frac{1}{7}$.
Furthermore, the outbreak size $\rho_F^{\infty}$ in layer $b$ exhibits a discontinuous jump to a complete breakdown ($\rho_F^{\infty} = 1$) when $\beta$ increases and surpasses another transition point $\beta_c^b \approx 0.159$. However, at the transition point $\beta_c^b$, only about 28.6\% of the population has been infected in layer $a$.

This example again indicates that the cascading failure propagation in layer $b$ can drastically amplify the impact of the epidemic outbreaks in layer $a$. Lastly, we remark that whether layer $b$ will exhibit a discontinuous transition or not depends on the values of $K$ and $\Theta$~\cite{Altarelli2013}, as predicted by the bootstrap percolation theory~\cite{Chalupa1979}.

\subsection*{Mitigation of Infection-induced Cascades}\label{sec:mitigation}

\subsubsection*{The Optimization Framework}\label{sec:opt_framework}
The catastrophic breakdown can be mitigated if timely protections are provided to stop the infection's spread. In our model, this is implemented by assigning a non-zero protection probability $\gamma_i(t)$ to node $i$, after which it is immune from infection from layer $a$.
To minimize the size of final failures, it would be more effective to take into account the spreading processes in both layers $a$ and $b$ when deciding which nodes to prioritize for protection.

Here, we develop mitigation strategies by solving the following 
constrained optimization problems
\begin{align}
& \min_{ \gamma } \;\; \mathcal{O}(\gamma) := \rho_F(T) = \frac{1}{N} \sum_{i \in V_b} P_F^i(T), \label{eq:objf_def} \\
& \text{s. t. } \;\; 0 \leq \gamma_i(t) \leq 1 \quad \forall i, t, \label{eq:constr_each_gamma}  \\
& \qquad \;\; \sum_{i \in V_b} \sum_{t = 0}^{T-1} \gamma_i(t) \leq \gamma^{\text{tot}}, \label{eq:constr_sum_gamma}
\end{align}
where the constraint in Eq.~(\ref{eq:constr_each_gamma}) ensures that $\gamma_i(t)$ is a probability, and Eq.~(\ref{eq:constr_sum_gamma}) represents the global budget constraint on the protection resources.
As the objective function $\mathcal{O}(\gamma)$ (the size of final failures) depends on the evolution of the two-layer spreading processes, the optimization problem is challenging. Lokhov and Saad introduced the optimal control framework to tackle similar problems, by estimating the marginal probabilities of individuals with the DMP methods~\cite{Lokhov2017}.
The success of the optimal control approach highlights another advantage of the theoretical methods over numerical simulations~\cite{Lokhov2017, HanlinSun2019, Zhou2022}.

In this work, we adopt a similar strategy to solve the optimization problem defined in Eqs.~(\ref{eq:objf_def})-(\ref{eq:constr_sum_gamma}), where $P_F^i(T)$ is estimated by the DMP equations derived in Methods. 
As the expressions of the DMP equations have been explicitly given and only involve elementary arithmetic operations, we leverage tools of automatic differentiation to compute the gradient of the objective function $\nabla_{\gamma} \mathcal{O}(\gamma)$ in a back-propagation fashion~\cite{Baydin2017}. It allows us to derive gradient-based algorithms for solving the optimization problem.
We remark that such a back-propagation algorithm is equivalent to optimal control with gradient descent update on the control parameters~\cite{QianxiaoLi2018}.
To save computing time, we adopt the approximation made in the subsection ``Simplification under Small Inter-layer Overlap'' in Methods for conducting the optimization; but we always use the full DMP formulations developed in the subsection ``Full Node-level DMP Equations'' in Methods for the evaluation of the outcomes.
This is particularly suitable for networks having little inter-layer overlaps. In scenarios where significant inter-layer overlaps exist and precision is crucial, it is always possible to resort to the complete version of the DMP equations.

To handle the box constraint in Eq.~(\ref{eq:constr_each_gamma}),
we adopt the mirror descent method, which performs the gradient-based update in the dual (or mirror) space rather than the primal space where $\{ \gamma_i(t) \}$ live~\cite{Nemirovski1983, Beck2003}.
In our case, we use the logit function $\Psi(x) = \log(\frac{x}{1-x})$ to map the primal control variable $\gamma_i(t)$ to the dual space as $h_i(t) = \psi( \gamma_i(t) ) \in \mathbb{R}$, where the gradient descent updates are performed.
The primal variable can be recovered through the inverse mapping of $\Psi(\cdot)$, which is $\Psi^{-1}(h) = \frac{1}{1+\exp(-h)}$.
The elementary mirror descent update step is
\begin{align}
g^n & \leftarrow \nabla_{\gamma} \mathcal{O}(\gamma^n), \label{eq:grad_original} \\
\gamma^{n+1} & \leftarrow \Psi^{-1} \big( \Psi(\gamma^{n}) - s g^n \big), \label{eq:gamma_mirror_update}
\end{align}
where $n$ is an index keeping track of the optimization process and $s$ is the step size of the gradient update.

In general, the above optimization process tends to increase the total resources $\sum_{i, t} \gamma_i(t)$. To prevent the violation of the constraint in Eq.~(\ref{eq:constr_sum_gamma}) during the updates, we suppress the gradient component which increases the total resources when $\sum_{i, t} \gamma_i(t) \geq (1 - \epsilon)  \gamma^{\text{tot}}$, by shifting the gradient $g^n$ in Eq.~(\ref{eq:gamma_mirror_update}) with a magnitude $b^n$
\begin{align}
b^n & \leftarrow \frac{\sum_{t,i}\gamma_i^n(t)(1-\gamma_i^n(t))\frac{\partial}{\partial\gamma_i(t)} \mathcal{O} (\gamma^{n})}{\sum_{t,i}\gamma_i^n(t)(1-\gamma_i^n(t))}, \\
g^n & \leftarrow \nabla_{\gamma} \mathcal{O}(\gamma^n) - b^n. \label{eq:grad_shifted}
\end{align}
The rationale for the choice of $b^n$ is explained in Supplementary Note 4. In our implementation of the algorithm, we choose $\epsilon = 0.02$.
Even though the shifted gradient method is used, it does not strictly forbid the violation of the constraint in Eq.~(\ref{eq:constr_sum_gamma}). If the resource capacity constraint is violated, we project the control variables to the feasible region through the simple rescaling
\begin{align}
\gamma^n \leftarrow \frac{\gamma^{\text{tot}}}{\sum_{t,i} \gamma^n_i(t)} \gamma^n.
\end{align}

Finally, the resource capacity constraint Eq.~(\ref{eq:constr_sum_gamma}) implies that a $\gamma^{\text{tot}}$ amount of protection resources can be distributed in different time steps. In some scenarios, the resources arrive in an online fashion, e.g., a limited number of vaccines can be produced every day. In these cases, there is a resource capacity constraint at each time step. Some results of such a scenario are discussed in Supplementary Note 5.

\subsubsection*{Case Study in a Tree Network}\label{sec:case_study_tree}
\begin{figure}
\hspace*{-1em}\includegraphics[scale=1.8]{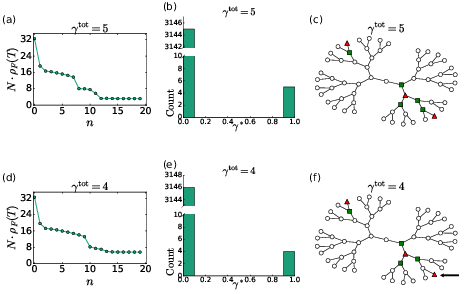}
\caption{Mitigation of the network failures in a binary tree network of size $N=63$, where both layers have the same topology. Panels (a)(b)(c) correspond to the case with $\gamma^{\text{tot}} = 5$, while Panels (d)(e)(f) correspond to the case with $\gamma^{\text{tot}} = 4$. 
Panels (a) and (d) depict how the final failure size changes during the optimization process. 
Specifically, the control parameters $\{ \gamma^n_i(t) \}$ for each optimization step $n$ were recorded, which were fed to the DMP equations for computing $\rho_F(T)$ at step $n$.
Panels (b) and (e) plot the histogram of the optimal decision variables $\{ \gamma^*_i(t) \}$. Panels (c) and (f) show the optimal placement of resources on layer $a$, where green square nodes receive protection (having a high $\gamma^*_i(t)$ at time $t=0$). The three red triangle nodes are the initially infected individuals. The system parameters are set as $\beta_{ji} = 0.5, \mu_i = 0.5, b_{ji} = 1, \Theta_i = 0.6 |\partial^b_i|$.}
\label{fig:control_case1}
\end{figure}

We first verify the effectiveness of the optimization method by considering a simple problem on a binary tree network of size $N = 63$, where both layers have the same topology.
The results are shown in Fig.~\ref{fig:control_case1}, where three individuals are chosen to be the infected seeds at time $t = 0$, and the outbreak is simulated for $T = 50$ time steps. 
Without any mitigation strategy, \emph{more than half of the population fail} at the end of the process.

We then protect some vital nodes to mitigate the system failure, by using the optimization method proposed above. In Fig.~\ref{fig:control_case1}(a)(b)(c), we restrict the total resources to be $\gamma^{\text{tot}} = 5$. Fig.~\ref{fig:control_case1}(a) shows that the optimization algorithm successfully reduces the final failure rate, which demonstrates the effectiveness of the method.
We found that the optimal protection resource distribution $\{ \gamma^*_i(t) \}$ mostly concentrates on a few nodes at a certain time step (as shown in Fig.~\ref{fig:control_case1}(b)), which implies that we can confidently select which nodes to protect. 
All the nodes with high $\gamma^*_i(t)$ receive protection at time $t=0$, which implies that the best mitigation strategy in this example is to distribute all $\gamma^{\text{tot}}$ resources as early as possible to stop the infection spread. Fig.~\ref{fig:control_case1}(c) shows the optimal placement of resources, which can completely block the infection spread, hence minimizing the network failure. In this example, both layers $a$ and $b$ have the same network structure, which is depicted in Fig.~\ref{fig:control_case1}(c).

Similar phenomena are observed in the case with $\gamma^{\text{tot}} = 4$ as shown in Fig.~\ref{fig:control_case1}(d)(e)(f), except that the protections are not sufficient to completely block the infection spread. The optimization algorithm sacrifices only two nodes in the vicinity of the infected node in the lower right corner of Fig.~\ref{fig:control_case1}(f) (indicated by a black arrow), leaving other parts of the network in the normal state. 

\onecolumngrid

\twocolumngrid

\begin{figure}
\centering
\includegraphics[scale=0.37]{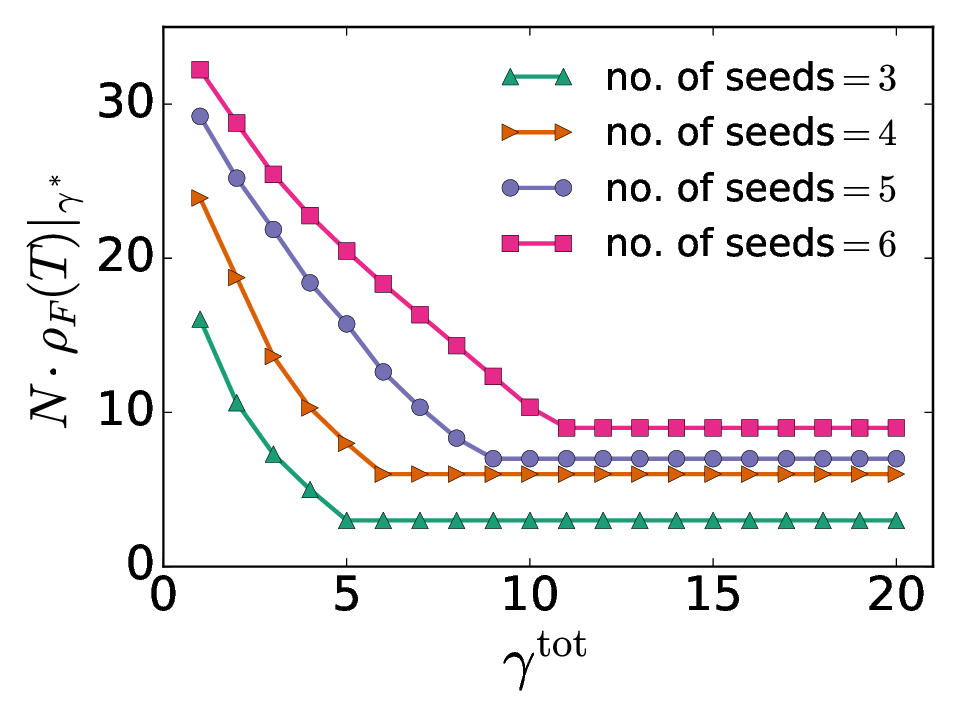}
\caption{Final failure size $N \cdot \rho_F(T)$ of a binary tree network evaluated at the optimal solution $\{ \gamma_i^*(t) \}$, as a function of the amount of total resources $\gamma^{\text{tot}}$. Different curves correspond to different number of initially infected seeds. 
The network topology and the system parameters are the same as those in Fig.~\ref{fig:control_case1}.
}
\label{fig:vary_gamma_tot}
\end{figure}

In Fig.~\ref{fig:vary_gamma_tot}, we further examine the influence of the total resource availability, i.e., $\gamma^{\text{tot}}$, on the final failure size $N \cdot \rho_F(T)$ determined at the optimal solution ${ \gamma_i^*(t) }$. It is observed that when $\gamma^{\text{tot}}$ increases, the failure size (at the optimum) firstly decreases monotonically, and then saturate when $\gamma^{\text{tot}}$ reaches a certain value such that there are enough protection resources to completely block the infection transmission. Another interesting observation is that for the cases with more initially infected seeds, introducing additional units of protection resource yields a less effective reduction in failure size compared to the cases with fewer initial infected seeds.

The good performance of the optimization is based on the fact that there are enough protection resources (i.e., having a large $\gamma^{\text{tot}}$) as well as being aware of the origins of the outbreak. In some cases, whether a node was infected at the initial time is not fully determined but follows a probability distribution. Such cases can be easily accommodated in the DMP framework which is intrinsically probabilistic. We investigated such a scenario with probabilistic seeding in Supplementary Note 6, and found that the optimization method can still effectively reduce the sizes of network failures.

\subsubsection*{Case Study in a Synthetic Network} \label{sec:case_study_synthetic}
To further showcase the applicability of the optimization algorithm for failure mitigation, we consider a synthetic technological multiplex network where layer $a$ represents a communication network and layer $b$ represents a power network.
We consider the scenario that the communication network can be attacked by malware but can also be protected by technicians, which is modeled by the proposed SIRP model. The infection of a node in the communication network causes the breakdown of the corresponding node in the power network. The breakdown of components in a power network can trigger further failures and form a cascade, which is modeled by the proposed LTM model. We have neglected the details of the power flow dynamics in order to obtain a tractable model and an insightful simple example.

\begin{figure}[bp!]
\includegraphics[scale=2.2, width=\textwidth]{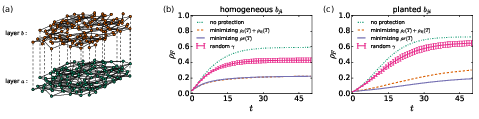}
\caption{A synthetic two-layer network and the evolution of its failure rate.
(a) The structure of the two-layer network, where each layer has $N=118$ nodes. Layer $a$ is a Watts-Strogatz small-world network, which mimics the topology of communication networks; it is obtained by rewiring a regular graph of degree $4$ with rewiring probability $p_\text{rewire} = 0.3$.
Layer $b$ is a power network extracted from the IEEE 118-bus test case.
(b) Evolution of the failure rate $\rho_F(t)$ under various mitigation strategies under homogeneous $\{ b_{ji} \}$. The curve labeled by ``random $\gamma$'' corresponds to the random deployment of a $\gamma^{\text{tot}}$ amount of protection resources at time $t=0$; 20 different random realizations are considered and the error bar indicates one standard deviation of the sample fluctuations. The time window is set as $T = 50$. Most system parameters are homogeneous with $\beta_{ji} = 0.2, \mu = 0.5, b_{ji} = 1$, while $\Theta_i = 0.6|\partial_i^b|$. Five nodes are randomly chosen as the in1itially infected individuals, and $\gamma^{\text{tot}} = 10$ is considered.
(c) Evolution of the failure rate $\rho_F(t)$ under various mitigation strategies under planted $\{ b_{ji} \}$. The system parameters are $\beta_{ji} = 0.17, \mu = 0.5, \Theta_i = 0.6|\partial_i^b|$. Planted influence parameters $\{ b_{ji} \}$ are considered. Three nodes are randomly chosen as the initially infected individuals, and $\gamma^{\text{tot}} = 9$ is considered. Other experiment set-ups are identical to those in Panel (b). 
} 
\label{fig:N118_net_and_control_case35}
\end{figure}

Here, we extract the network topology from the IEEE 118-bus test case to form layer $b$~\cite{Christie1993}, which has $N = 118$ nodes.
We then obtain layer $a$ by rewiring a regular graph of the same size with degree $K = 4$ using a rewiring probability $p_\text{rewire} = 0.3$, which creates a Watts-Strogatz small-world network and mimics the topology of communication networks~\cite{Cai2017}. The resulting multiplex network is plotted in Fig.~\ref{fig:N118_net_and_control_case35}(a).

As the failures in layer $b$ are initially induced by the infections in layer $a$, one may wonder whether deploying the protection resources by minimizing the size of infections, i.e., minimizing $\rho_I(T) + \rho_R(T)$ instead of minimizing $\rho_F(T)$, is already sufficient to mitigate the final failures. 
To investigate this effect, we replace the objective function in Eq.~(\ref{eq:objf_def}) by $\mathcal{O}^a (\gamma) = \rho_I(T) + \rho_R(T)$ and solve the optimization problem using the same techniques in the subsection ``The Optimization Framework''. The result is shown in Fig.~\ref{fig:N118_net_and_control_case35}(b), which suggests that blocking the infection is as good as minimizing the original objective function in Eq.~(\ref{eq:objf_def}) for the purpose of minimizing the total failure size. Minimizing either objective function constitutes a much better improvement over the random deployment of the same amount of protection resources in this case.

The results in Fig.~\ref{fig:N118_net_and_control_case35}(b) point to the conventional wisdom that one should try best to stop the epidemic or malware spread (in layer $a$) for mitigating system failure. The situation will be different if there are vital components in layer $b$, which should be protected to prevent the failure cascade. This is typically manifested in the heterogeneity of the network connectivity or the system parameters. 
To showcase this effect, we manually plant a vulnerable connected cluster in layer $b$ by setting the influence parameters $b_{ji}$ for an edge $(i,j)$ in this cluster as $b_{ji} > \Theta_i$, so that the failure of node $j$ itself is already sufficient to trigger the failure of node $i$.
Such a set-up is relevant for commercial, industrial and engineering networks, among others; e.g., supply chain networks evolve to enhance their throughput and efficiency but may operate with little redundancy and low robustness.
In this case, we found that minimizing $\rho_F(T)$ yields a much better improvement over minimizing $\rho_I(T) + \rho_R(T)$ for the purpose of mitigating the system failure, as shown in Fig.~\ref{fig:N118_net_and_control_case35}(c).

\onecolumngrid

\twocolumngrid

\phantom{.}
\phantom{.}
\phantom{.}

\subsubsection*{Case Studies in a Real-world Social Networks}\label{sec:case_study_tailor_shop}
\begin{figure}
\centering
\includegraphics[scale=1.05]{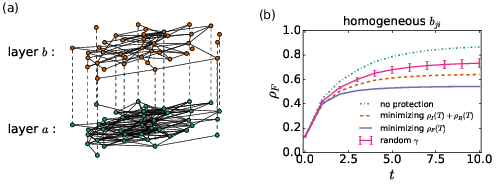}
\caption{The Kapferer's tailor shop network and the evolution of its failure rate.
(a) The structure of the Kapferer's tailor shop network, which involves interactions among 39 workers in a tailor shop in Zambia during a period of one month. Layer $a$ represents the ``sociational'' relations, while layer $b$ represents the ``instrumental'' relations. (b) Evolution of the failure rate $\rho_F(t)$ of the tailor shop network up to time $T = 10$ under various mitigation strategies.
The ``random $\gamma$'' strategy has the same set-up as the one in Fig.~\ref{fig:N118_net_and_control_case35}(b); 20 different random realizations are considered and the
error bar indicates one standard deviation of the sample fluctuations..
Most system parameters are homogeneous with $\beta_{ji} = 0.07, \mu = 0.5, b_{ji} = 1$, while $\Theta_i = 0.6|\partial_i^b|$. Five nodes with the highest degrees in layer $a$ are selected as the initially infected individuals, and $\gamma^{\text{tot}} = 10$ is considered. 
}
\label{fig:tailor_shop_net_and_control}
\end{figure}

Lastly, we examine the Kapferer's tailor shop network, a well-known social network dataset gathered by B. Kapferer in Zambia, documenting interactions among workers in a tailor shop~\cite{Kapferer1972, KapfererUCI}. 
This dataset records two types of interactions across two different time frames. The first interaction type is termed ``sociational'', which encapsulates friendship and socioemotional relationship among the workers. The second interaction type is termed ``instrumental'', which reflects work- and assistance-related connections among them. 
For our analysis, the ``sociational'' network observed in the initial time frame is assigned to layer $a$, acting as the substrate for infection transmission, while the corresponding ``instrumental'' network is assigned to layer $b$, where the failure (in terms of work accomplishment) of a node can be triggered by the malfunctioning of its neighboring nodes.
These networks are treated as undirected graphs for simplicity. The resulting two-layer network is depicted in Fig.~\ref{fig:tailor_shop_net_and_control}(a).

We assign homogeneous values to the majority of system parameters without deliberately introducing any vulnerable component in the network; the set-up closely aligns with the scenario depicted in Fig.~\ref{fig:N118_net_and_control_case35}(b) and presents a stark contrast to the scenario in Fig.~\ref{fig:N118_net_and_control_case35}(c).
We select five nodes that possess the highest degrees to serve as the initially infected individuals, which can be viewed as super-spreaders in the network. We then protect the vital nodes to mitigate the system failures by using the optimization method as above, where the result is shown in Fig.~\ref{fig:tailor_shop_net_and_control}(b). 
Interestingly, minimizing the size of failures (i.e., $\rho_F(T)$) is evidently better than minimizing the size of infections (i.e., $\rho_I(T) + \rho_F(T)$) for the purpose of failure mitigation. It suggests that in this realistic and natural scenario, simply blocking the infection transmission is sub-optimal and one needs to take a holistic view of the two-layer model for optimizing the network's utility.

\section*{Conclusion}\label{sec:conclusion}
We investigate the nature of a type of  two-layer spreading processes in unidirectionally dependent networks, comprising two interacting layers $a$ and $b$. Disease or malware spreads in layer $a$, which can trigger cascading failures in layer $b$, leading to secondary disasters.
The spreading processes in the two layers are modeled by the SIRP and LTM models, respectively. To tackle the complex stochastic dynamics in the two-layer networks, we utilized the dynamic message-passing method by working out the dynamic belief propagation equations. The resulting DMP algorithms have low computational complexity in sparse networks and allow us to perform accurate and efficient inference of the system states.

Based on the DMP method, we systematically studied and evaluated the impact of the infection-induced cascading failures. The cascade process in layer $b$ can lead to large-scale network failures, even when the infection rate in layer $a$ remains at a relatively low level. 
By considering a homogeneous network topology and homogeneous system parameters, we derive the asymptotic and large-size limits of the DMP equations. The asymptotic limit of the two-layer spreading processes corresponds to the coupling between a bond percolation model and a bootstrap percolation model, which can be analytically solved. 
The infection outbreak size in layer $a$ changes continuously from zero to non-zero as the infection probability $\beta$ surpasses a transition point $\beta^a_c$, while the failure size in layer $b$ can exhibit a discontinuous jump to the completely failed state when $\beta$ surpasses another transition point $\beta^b_c$ under certain conditions.
All these results highlight the observation that cascading failure propagation in layer $b$ can drastically amplify the impact of the epidemic outbreaks in layer $a$, which requires special attention.

Another advantage of the DMP method is that it yields a set of closed-form equations, which can be very useful for other downstream analyses and tasks. We exploited this property to devise optimization algorithms for mitigating network failure. The optimization method works by back-propagating the impact at the final time to adjust the control parameters (i.e., the protection probabilities). The mirror descent method and a heuristic gradient shift method were also used to handle the constraints on the control parameter. We show that the resulting algorithm can effectively minimize the size of system failures. We believe that our dedicated analyses provide valuable insights and a deeper understanding of the impact the infection-induced cascading failures on networks, and the obtained optimization algorithms will be useful for practical applications in systems of this kind.

\section*{Data Availability}
Datasets cited in this study are publicly accessible and have been referenced accordingly in the manuscript.

\section*{Code Availability}
Source codes of the methods and analyses used in this study are available at \url{https://github.com/boli8/DMP-for-SIRP-LTM}.

\begin{acknowledgments}
B.L. and D.S. acknowledge support from European Union's Horizon 2020 research and innovation programme under the Marie Sk{\l}odowska-Curie Grant Agreement No. 835913.
B.L. acknowledges support from the National Natural Science Foundation of China (Grant No. 12205066), the Shenzhen Start-Up Research Funds (Grant No. BL20230925) and the start-up funding from Harbin Institute of Technology, Shenzhen (Grant No. 20210134).
D.S. acknowledges support from the Leverhulme Trust (RPG-2018-092) and the EPSRC programme grant TRANSNET (EP/R035342/1).
\end{acknowledgments}

\section*{Author contributions}
B.L. and D.S. conceived the project and developed the theoretical framework. B.L. carried out the theoretical calculations and the numerical simulations. B.L. and D.S. discussed the results and prepared the manuscript.

\section*{Competing interests}
The authors declare no competing interests.

\section*{References}
\bibliography{reference}

\end{document}


\title{Supplementary Information for \\ ``Infection-induced Cascading Failures -- Impact and Mitigation''}

\author{Bo Li}
\email{libo2021@hit.edu.cn}
\affiliation{School of Science, Harbin Institute of Technology (Shenzhen), Shenzhen, 518055, China}
\affiliation{Non-linearity and Complexity Research Group, Aston University, Birmingham,
B4 7ET, United Kingdom}

\author{David Saad}
\email{d.saad@aston.ac.uk}
\affiliation{Non-linearity and Complexity Research Group, Aston University, Birmingham,
B4 7ET, United Kingdom}

\maketitle

\section*{Supplementary Note 1: Deriving the DMP Equations From Dynamic Belief Propagation}\label{app:deriveDMP}
In this section, we supplement some technical details of the DMP equations based on dynamic belief propagation. 

Firstly, we provide a summary of the main notations used throughout the study in Supplementary Table.~\ref{tab:notations}.

\renewcommand{\arraystretch}{1.15} 

\begin{table}[ht]
\centering
\begin{tabular}{p{3cm} | p{13cm}}
\hline
\textbf{Symbol} & \textbf{Description} \\
\hline
$ S, I, R, P $       & susceptible, infected, recovered, protected  (i.e., the four possible states in layer $a$) \\
\hline
$ N, F $             & normal, failed (i.e., the two possible states in layer $b$) \\
\hline
$ x_i  $             & $x_i = 0$ denotes that node $i$ is in state $N$, while $x_i = 1$ denotes that node $i$ is in state $F$  \\
\hline
$ \beta_{ji} $       & the probability of infection transmission from node $j$ (in state $I$) to node $i$ (in state $S$) \\
\hline
$ \gamma_i(t) $      & the probability that a susceptible node $i$ turns into the protected state at time $t + 1$ \\
\hline
$ \Theta_i $         & a threshold parameter of node $i$ below which the node fails \\
\hline
$ b_{ji} $           & the influence parameter which measures the importance of the failure of node $j$ on node $i$ \\
\hline
$ \partial_i^a $     & the set of nodes adjacent to node $i$ in layer $a$ \\
\hline
$ \partial_i^b $     & the set of nodes adjacent to node $i$ in layer $b$ \\
\hline
$ \partial_i $       & $ \partial_i^a \cup \partial_i^b $, i.e., the set of nodes adjacent to node $i$ in the combined network of layers $a$ and $b$ \\
\hline
$ \tau_i^a $         & the first time at which node $i$ turns into state $I$ \\
\hline
$ \omega_i^a $       & the first time at which node $i$ turns into state $R$ \\
\hline
$ \varepsilon_i^a $  & the first time at which node $i$ turns into state $P$ \\
\hline
$ \tau_i^b $         & the first time at which node $i$ turns into state $F$ \\
\hline
$ m^{i}(\tau_{i}^{a},\omega_{i}^{a},\varepsilon_{i}^{a},\tau_{i}^{b}) $    & the probability that node $i$ follows a trajectory parametrized by its state-transition times $(\tau_{i}^{a},\omega_{i}^{a},\varepsilon_{i}^{a},\tau_{i}^{b})$ \\
\hline
$ m^{i \to j}(\tau_{i}^{a},\omega_{i}^{a},\varepsilon_{i}^{a},\tau_{i}^{b}) $    & similar to $ m^{i}(\tau_{i}^{a},\omega_{i}^{a},\varepsilon_{i}^{a},\tau_{i}^{b}) $ except that it is defined in the cavity graph where node $j$ is removed \\
\hline
$ P_S^{i}(t) $       & the probability that node $i$ is in state $S$ at time $t$ \\
\hline
$ P_S^{i \to j}(t) $ & similar to $P_S^{i}(t)$ except that it is defined in the cavity graph where node $j$ is removed \\
\hline
$ \rho_F(t) $        & the fraction of nodes that have failed at time $t$ \\
\hline
$ \mathbb{I}(y) $    & indicator function which returns $1$ if $y$ is true, and returns $0$ if $y$ is false \\
\hline
$ \delta_{y, z} $    & the Kronecker delta function \\
\hline
\end{tabular}
\caption{Summary of the main notations.}
\label{tab:notations}
\end{table}

We assume that at the initial time $t = 0$, node $i$ is either in state $S$ or state $I$, occurring with probabilities $P_S^i(0)$ and $P_I^i(0)$ (with $P_S^i(0) + P_I^i(0) = 1$), respectively. 

According to dynamical rule of the SIRP model defined in the main text, the transition kernel $W_{\text{SIRP}}^{i}(\cdot)$ of the spreading process in layer $a$ admits the following form
\begin{align}
& W_{\text{SIRP}}^{i}(\tau_{i}^{a},\omega_{i}^{a},\varepsilon_{i}^{a}||\{\tau_{k}^{a},\omega_{k}^{a},\varepsilon_{k}^{a}\}_{k\in\partial_{i}^{a}}) \\
= & \;\; \mathbb{I}(\tau_i^a < \varepsilon_i^a) \bigg\{ P_I^i(0) \mathbb{I} ( \tau_i^a = 0 ) + P_S^i(0) \mathbb{I} ( \tau_i^a > 0 ) \prod_{t' = 0}^{\tau_i^a - 2} \prod_{k \in \partial_i^a} \mathbb{I} \big[ 1 - \beta_{ki} \mathbb{I}( \omega_k^a \geq t'+1 ) \mathbb{I}( \tau_k^a \leq t' ) \big] \nonumber \\
& \qquad \qquad \qquad \qquad  \times \bigg[ 1 - \prod_{k \in \partial_i^a} \big[ 1 - \beta_{ki} \mathbb{I}(\omega_k^a \geq \tau_i^a) \mathbb{I}( \tau_k^a \leq \tau_i^a - 1 ) \big] \bigg]  \times \bigg( \prod_{t'' = \tau_i^a}^{\omega_i^a - 2} (1 - \mu_i) \bigg) \mu_i \times \prod_{t''' = 0}^{\tau_i^a - 1} ( 1 - \gamma_i(t''')) \bigg\} \nonumber \\
& \;\; + \mathbb{I}( \tau_i^a \geq \varepsilon_i^a ) \bigg\{ P_S^i(0) \mathbb{I} ( \tau_i^a > 0 ) \prod_{t' = 0}^{\varepsilon_i^a - 2} \prod_{k \in \partial_i^a} \mathbb{I} \big[ 1 - \beta_{ki} \mathbb{I}( \omega_k^a \geq t'+1 ) \mathbb{I}( \tau_k^a \leq t' ) \big] \times \big[ \prod_{t''' = 0}^{\varepsilon_i^a - 2} ( 1 - \gamma_i(t''')) \big] \gamma_i(\varepsilon_i^a - 1)   \bigg\}. \nonumber
\end{align}
The transition kernel $W_{\text{LTM}}^{i}(\cdot)$ of the cascade process in layer $b$ admits the following form
\begin{align}
& W_{\text{LTM}}^{i}( \tau_{i}^{b}||\tau_{i}^{a},\varepsilon_{i}^{a},\{\tau_{k}^{b}\}_{k\in\partial_{i}^{b}} ) \\
= & \;\; \mathbb{I}\bigg[\sum_{k\in\partial_{i}^{b}}b_{ki}\mathbb{I}(\tau_{k}^{b}\leq\tau_{i}^{b}-2)< \threshold_i \bigg]\delta_{\tau_{i}^{b},\tau_{i}^{a}}  + \mathbb{I}(\tau_{i}^{b}<\tau_{i}^{a})\mathbb{I}\bigg[\sum_{k\in\partial_{i}^{b}}b_{ki}\mathbb{I}(\tau_{k}^{b}\leq\tau_{i}^{b}-2)< \threshold_i \bigg] \mathbb{I}\bigg[\sum_{k\in\partial_{i}^{b}}b_{ki}\mathbb{I}(\tau_{k}^{b}\leq\tau_{i}^{b}-1)\geq \threshold_i \bigg], \nonumber
\end{align}
where the first term corresponds to the case where node $i$ fails (in layer $b$) due to infection (from layer $a$), while the second term corresponds to the case where node $i$ failed due to losing supports from neighboring nodes $\partial_i^b$ in layer $b$.

For infection spread in layer $a$, the probability of node $i$ in a certain state is computed by tracing over the corresponding probabilities of trajectories $m_{a}^{i \to j}( \cdot )$ (note that the process in layer $b$ does not exert any influence on layer $a$)
\begin{align}
& P_{S}^{i\to j}(t) = \sum_{\tau_{i}^{a}, \omega_{i}^{a}, \varepsilon_{i}^{a}} m_a^{i\to j} (\tau_{i}^{a},\omega_{i}^{a},\varepsilon_{i}^{a}) \; \mathbb{I}(t<\tau_{i}^{a}<\omega_{i}^{a}) \; \mathbb{I}( t <\varepsilon_{i}^{a} ), \\
& P_{I}^{i\to j}(t) = \sum_{\tau_{i}^{a}, \omega_{i}^{a}, \varepsilon_{i}^{a} } m_a^{i\to j} (\tau_{i}^{a},\omega_{i}^{a},\varepsilon_{i}^{a} ) \; \mathbb{I}(\tau_{i}^{a} \leq t < \omega_{i}^{a}) \; \delta_{\varepsilon_i^a, \infty},  \\
& P_{R}^{i\to j}(t) = \sum_{\tau_{i}^{a}, \omega_{i}^{a}, \varepsilon_{i}^{a} } m_a^{i\to j} (\tau_{i}^{a},\omega_{i}^{a},\varepsilon_{i}^{a} ) \; \mathbb{I}(\tau_{i}^{a} < \omega_{i}^{a} \leq t) \; \delta_{\varepsilon_i^a, \infty},  \\
& P_{P}^{i\to j}(t) = \sum_{\tau_{i}^{a}, \omega_{i}^{a}, \varepsilon_{i}^{a} } m_a^{i\to j} (\tau_{i}^{a},\omega_{i}^{a},\varepsilon_{i}^{a} ) \; \mathbb{I}( \varepsilon_i^a \leq t) \; \delta_{\tau_i^a, \infty}.
\end{align}
The computation of these probabilities is similar to that of the SIR model, where we refer readers to Ref.~\cite{Lokhov2015} for the details.

For activities in layer $b$, the probability of node $i$ in state $F$ (failed) is computed as
\begin{align}
P_F^{i \to j}(t) & = \sum_{\tau_{i}^{a}, \omega_{i}^{a}, \varepsilon_{i}^{a}, \tau_{i}^{b} }  \mathbb{I}(\tau_{i}^{b} \leq t) m^{i\to j} (\tau_{i}^{a},\omega_{i}^{a},\varepsilon_{i}^{a}, \tau_{i}^{b}) \nonumber \\
& = \sum_{\tau_{i}^{a}, \omega_{i}^{a}, \varepsilon_{i}^{a}, \tau_{i}^{b} }  \mathbb{I}(\tau_{i}^{b} \leq t) m_a^{i\to j} (\tau_{i}^{a},\omega_{i}^{a},\varepsilon_{i}^{a} ) m_b^{i\to j} ( \tau_{i}^{b} \mid \tau_{i}^{a},\varepsilon_{i}^{a} )
\end{align}
which appears much more difficult to treat due to the dependence on the activities in layer $a$. In particular, the failure of node $i$ can be attributed to the infection from one of its neighbors from layer $a$, or to the failures of its neighbors from layer $b$. For the latter case, node $i$ can be either in state $S$ or in state $P$ at time $t$, which depends on the infection time $\tau_i^a$ and protection time $\varepsilon_i^a$. 
For this reason, we introduce the following conditional failure probability
\begin{align}
P_F^{i\to j}(t | \tau_{i}^{a}, \omega_{i}^{a}, \varepsilon_{i}^{a}) & = \sum_{\tau_{i}^{b}} \mathbb{I}(\tau_{i}^{b} \leq t) m_{b}^{i\to j}(\tau_{i}^{b}|\tau_{i}^{a},\varepsilon_{i}^{a}) = \sum_{\tau_{i}^{b}} \mathbb{I}(\tau_{i}^{b}\leq t)\frac{m^{i\to j}(\tau_{i}^{a},\omega_{i}^{a},\varepsilon_{i}^{a},\tau_{i}^{b})}{m_{a}^{i\to j}(\tau_{i}^{a},\omega_{i}^{a},\varepsilon_{i}^{a})} \nonumber \\
& =\mathbb{I}(\tau_{i}^{a}\leq t) + \mathbb{I}(\tau_i^a > t) \frac{ \xi^{i\to j}(t | \tau_{i}^{a}, \omega_{i}^{a}, \varepsilon_{i}^{a}) }{ m_{a}^{i\to j}(\tau_{i}^{a},\omega_{i}^{a},\varepsilon_{i}^{a}) } ,
\end{align}
where we have defined
\begin{align}
\xi^{i\to j}(t|\tau_{i}^{a}, \omega_{i}^{a}, \varepsilon_{i}^{a}) := \mathbb{I}(\tau_i^a > t)  \sum_{\tau_{i}^{b}} \mathbb{I}(\tau_{i}^{b} \leq t) m^{i\to j}(\tau_{i}^{a},\omega_{i}^{a},\varepsilon_{i}^{a},\tau_{i}^{b} ),
\end{align}
which is the cavity probability of node $i$ not in state $I$ or $R$ but being failed at time $t$ due to the failures of neighbors from layer $b$, while it follows the specific trajectory $\{ \tau_{i}^{a}, \omega_{i}^{a}, \varepsilon_{k}^{a} \}$ in layer $a$.

The marginal probability of node $i$ in state $F$ at time $t$ is obtained by tracing over all the possible trajectories of layer $a$ as
\begin{align}
P_F^{i \to j}(t) = & \sum_{\tau_{i}^{a},\omega_{i}^{a},\varepsilon_{i}^{a}}\mathbb{I}(\omega_{i}^{a}>\tau_{i}^{a}) m_{a}^{i\to j}(\tau_{i}^{a},\omega_{i}^{a},\varepsilon_{i}^{a}) P_F^{i \to j}(t | \tau_{i}^{a}, \omega_{i}^{a}, \varepsilon_{i}^{a})\nonumber \\
= & P_{I}^{i \to j}(t) + P_{R}^{i \to j}(t) + \sum_{\tau_{i}^{a},\omega_{i}^{a},\varepsilon_{i}^{a}} \mathbb{I}(\tau_{i}^{a}>t) \mathbb{I}(\omega_{i}^{a}>\tau_{i}^{a}) \xi^{i \to j}(t | \tau_{i}^{a}, \omega_{i}^{a}, \varepsilon_{i}^{a}),
\end{align}
where the summation in the last term can be further decomposed into $P_{SF}^{i \to j}(t)$ and $P_{PF}^{i \to j}(t)$, depending on whether the protection on node $i$ (given at time $\varepsilon_i^a$) occurs after time $t$ or before time $t$
\begin{align}
& P_{SF}^{i \to j}(t) = \sum_{\tau_{i}^{a},\omega_{i}^{a},\varepsilon_{i}^{a}} \mathbb{I} ( \varepsilon_i^a > t ) \mathbb{I}(\tau_{i}^{a}>t) \mathbb{I}(\omega_{i}^{a}>\tau_{i}^{a}) \xi^{i \to j}(t | \tau_{i}^{a}, \omega_{i}^{a}, \varepsilon_{i}^{a}), \\
& P_{PF}^{i \to j}(t) = \sum_{\tau_{i}^{a},\omega_{i}^{a},\varepsilon_{i}^{a}} \mathbb{I} ( \varepsilon_i^a \leq t) \mathbb{I}(\tau_{i}^{a}>t) \mathbb{I}(\omega_{i}^{a}>\tau_{i}^{a}) \xi^{i \to j}(t | \tau_{i}^{a}, \omega_{i}^{a}, \varepsilon_{i}^{a}).
\end{align}
In summary, we can decompose $P_F^{i \to j}(t)$ into four terms
\begin{align}
P_F^{i \to j}(t) = P_{I}^{i \to j}(t) + P_{R}^{i \to j}(t) + P_{SF}^{i \to j}(t) + P_{PF}^{i \to j}(t),
\end{align}
where a similar form holds for $P_F^i(t)$ as stated in the main text.

\begin{figure}
\centering
\hspace{-1cm}\includegraphics[scale=2.6]{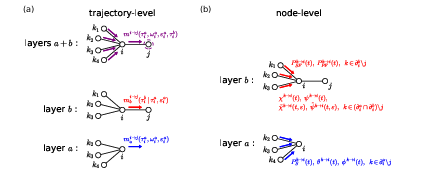}
\caption{Messages employed for the two-layer processes considered. (a) The trajectory-level messages $\{ m^{i \to j}(\cdot) \}$ live in the entire network comprising layers $a$ and $b$, which implies that $\{ m_a^{i \to j}(\cdot), m_b^{i \to j}(\cdot) \}$ are also defined on the entire network. (b) The messages $\{ P_{S}^{k\to i}(t), \theta^{k\to i}(t), \phi^{k\to i}(t) \}$ are only needed for edges belonging to layer $a$ where the SIRP model is defined. The node-level messages for the SIRP process only enter into the DMP equations for the LTM process through the overlapping neighbors $\partial_i^a \cap \partial_i^b$ (i.e., $\{ k_2, k_3 \}$ in this example).}
\label{fig:cav_graph_prob_demo}
\end{figure}

To obtain node-level iteration equations of $P_{SF}^{i \to j}(t)$ and $P_{PF}^{i \to j}(t)$, the key is to further introduce the auxiliary probabilities $\chi^{k \to i}(t)$, $\psi^{k \to i}(t)$, $\tilde{\chi}^{k \to i}(t, \varepsilon)$ and $\tilde{\psi}^{k \to i}(t, \varepsilon)$, defined as
\begin{align}
\chi^{k\to i}(t)= & \sum_{\tau_{k}^{a},\omega_{k}^{a},\varepsilon_{k}^{a}}\mathbb{I}(\omega_{k}^{a}>\tau_{k}^{a})\prod_{t'=0}^{t-1}\big[1-\beta_{ki}\mathbb{I}(\omega_{k}^{a}\geq t'+1)\mathbb{I}(\tau_{k}^{a}\leq t')\big]  m_{a}^{k\to i}(\tau_{k}^{a},\omega_{k}^{a},\text{\ensuremath{\varepsilon_{k}^{a}}}) P_F^{k\to i}(t-1|\tau_{k}^{a}, \omega_{k}^{a}, \varepsilon_{k}^{a}), \\
\psi^{k\to i}(t)= & \sum_{\tau_{k}^{a},\omega_{k}^{a},\varepsilon_{k}^{a}}\mathbb{I}(\omega_{k}^{a}>\tau_{k}^{a})\prod_{t'=0}^{t-1}\big[1-\beta_{ki}\mathbb{I}(\omega_{k}^{a}\geq t'+1)\mathbb{I}(\tau_{k}^{a}\leq t')\big]m_{a}^{k\to i}(\tau_{k}^{a},\omega_{k}^{a},\varepsilon_{k}^{a})\mathbb{I}(\tau_{k}^{a}\leq t-1), \\
\tilde{\chi}^{k\to i}(t,\varepsilon)= & \sum_{\tau_{k}^{a},\omega_{k}^{a},\varepsilon_{k}^{a}} \mathbb{I}(\omega_{k}^{a}>\tau_{k}^{a}) \prod_{t'=0}^{\varepsilon-2}\big[1-\beta_{ki}\mathbb{I}(\omega_{k}^{a}\geq t'+1)\mathbb{I}(\tau_{k}^{a}\leq t')\big]m_{a}^{k\to i}(\tau_{k}^{a},\omega_{k}^{a},\varepsilon_{k}^{a}) P_F^{k\to i}(t-1|\tau_{k}^{a},\omega_{k}^{a},\varepsilon_{k}^{a}), \\
\tilde{\psi}^{k\to i}(t,\varepsilon)= & \sum_{\tau_{k}^{a},\omega_{k}^{a},\varepsilon_{k}^{a}} \mathbb{I}(\omega_{k}^{a}>\tau_{k}^{a}) \prod_{t'=0}^{\varepsilon-2}\big[1-\beta_{ki}\mathbb{I}(\omega_{k}^{a}\geq t'+1)\mathbb{I}(\tau_{k}^{a}\leq t')\big]m_{a}^{k\to i}(\tau_{k}^{a},\omega_{k}^{a},\varepsilon_{k}^{a}) \mathbb{I}(\tau_{k}^{a}\leq t-1).
\end{align}
These auxiliary probabilities are linked to $P_{SF}^{i \to j}(t)$ and $P_{PF}^{i \to j}(t)$ through the transition kernel $W^i_{\text{SIRP}}$ and $W^i_{\text{LTM}}$, respectively, and their iteration equations can be mechanistically derived (e.g., by relating $\chi^{k\to i}(t)$ to $\chi^{k\to i}(t-1)$). The explicit forms of the iteration equations and the physical interpretation of the auxiliary probabilities are stated in the main text.
The trajectory-level and node-level messages employed for the two-layer processes are summarized in Fig.~\ref{fig:cav_graph_prob_demo}.

\section*{Supplementary Note 2: Testing the Non-overlapping Approximation}\label{app:testing_DMP_approx}
\begin{figure}
\centering
\includegraphics[scale=2]{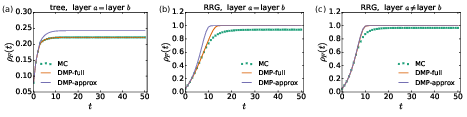}
\caption{Evolution of the failure rate $\rho_F(t)$ for different approaches. Here, ``DMP-full'' stands for the complete DMP formulations, ``DMP-approx'' stands for DMP with non-overlapping approximations, and ``MC'' stands for Monte Carlo simulations. The system parameters are $T = 50, \beta_{ji} = 0.2, \mu_i = 0.5, b_{ji} = 1, \Theta_i =  0.6 |\partial_i^b|, \gamma_i(t) = 0$. (a) Both layers are the same binary tree of size $N = 63$. (b) Both layers are the same RRG of size $N=100$ and degree $K=5$. (c) Both layers are RRG of size $N=100$ and degree $K=5$, but layers $a$ and $b$ have distinct topologies.}
\label{fig:rhoF_vs_t_dmp_mc}
\end{figure}

In this section, we test the efficacy of the non-overlapping approximation applied to the DMP equations in layer $b$ as introduced in the subsection ``Simplification under Small Inter-layer Overlap'' in the Methods section of the main text.
In particular, we compare the evolution of the expected cascade sizes $\rho_F(t) = 1/N \sum_{i \in V_b} P^i_F(T)$ for various approaches; the results are shown in Fig.~\ref{fig:rhoF_vs_t_dmp_mc}.
In Fig.~\ref{fig:rhoF_vs_t_dmp_mc}(c), both layers are random regular graphs (RRG) which are constructed independently, having distinct network topologies. The estimations given by the DMP method under the non-overlapping approximation matches with those from the full DMP method, as expected.
In Fig.~\ref{fig:rhoF_vs_t_dmp_mc}(a) and (b), both layers have the same network topology, therefore having the maximal neighbor overlap and violating the non-overlapping assumption. In these cases, the non-overlapping approximation for the DMP equations yields less accurate results compared to the complete treatment. Nonetheless, it still maintains the fundamental qualitative characteristics of the cascade dynamics.

\section*{Supplementary Note 3: Deriving the Large-time Limit of the Discrete-time SIR Model}
\label{app:SIR_asymptotic}
Here, we derive the DMP equations of the discrete-time SIR model in the large-time limit, which differs from the continuous-time counterpart~\cite{Karrer2010}.

By setting $\gamma_i(t) = 0$ in the DMP equations of the SIRP model provided in the main text, we recover the DMP equations for the traditional SIR model, where $\theta^{i \to j}$ and $\phi^{i \to j}$ admit the following expressions
\begin{align}
\theta^{i \to j}(t+1) - \theta^{i \to j}(t) = & -\beta_{ij} \phi^{i \to j}(t), \\
\phi^{i \to j}(t+1) - \phi^{i \to j}(t) = & -(\beta_{ij} + \mu_i - \beta_{ij}\mu_i) \phi^{i \to j}(t)   - \big[ P_S^{i \to j}(t+1) - P_S^{i \to j}(t) \big].
\end{align}
Summing both sides of the above equations from $t=0$ to $t=T-1$ and canceling the term $\sum_{t=0}^{T-1} \phi^{i \to j}(t)$ yields
\begin{align}
\phi^{i \to j}(T) - \phi^{i \to j}(0) = \frac{\beta_{ij} + \mu_i - \beta_{ij}\mu_i}{\beta_{ij}} \big[ \theta^{i \to j}(T) - \theta^{i \to j}(0) \big] - \big[ P_S^{i \to j}(T) - P_S^{i \to j}(0) \big].
\end{align}
Define $p_{ij} = \beta_{ij} / (\beta_{ij} + \mu_i - \beta_{ij} \mu_i)$ and remind that the initial conditions for the messages are $P_S^{i \to j}(0) = P_S^i(0), \phi^{i \to j}(0) = 1 - P_S^i(0), \theta^{i\to j}(0) = 1$, which leads to
\begin{align}
\theta^{i \to j}(T) = 1 - p_{ij} + p_{ij} P_S^{i \to j}(T) + p_{ij} \phi^{i \to j}(T).
\end{align}

When $T \to \infty$, all infected nodes will recover, which implies that $\phi^{i \to j}(\infty) = 0$ and $\theta^{i \to j}(\infty)$ admits the following expression 
\begin{align}
\theta^{i \to j}(\infty) & = 1 - p_{ij} + p_{ij} P_S^{i \to j}(\infty). \label{eq:theta_i_to_j_inf}
\end{align}
Recall that when $\gamma_i(t) = 0$, we have
\begin{align}
& P_S^{i \to j}(\infty) = P_S^{i}(0) \prod_{k \in \partial^a_i \backslash j } \theta^{k \to i}(\infty),  \label{eq:PS_i_to_j_inf} \\
& P_S^{i}(\infty) = P_S^{i}(0) \prod_{k \in \partial^a_i } \theta^{k \to i}(\infty), 
\end{align}
which leads to the self-consistent equations for the messages $\{ \theta^{i \to j}(\infty) \}$ as
\begin{align}
\theta^{i \to j}(\infty) = 1 - p_{ij} + p_{ij} P_S^{i}(0) \prod_{k \in \partial^a_i \backslash j } \theta^{k \to i}(\infty).
\end{align}

The large-time behavior of the discrete-time SIR model can be mapped to a bond percolation problem, as stated in the main text. This can be made transparent by introducing the probability that node $i$ has been infected in the cavity graph where node $j$ is absent as
\begin{align}
r_{i \to j} = 1 - P_S^{i \to j}(\infty). 
\end{align}
Substituting $P_S^{i \to j}(\infty) = 1 - r_{i \to j}$ into Eq.~(\ref{eq:theta_i_to_j_inf}) and Eq.~\ref{eq:PS_i_to_j_inf} leads to a recursive equations for the messages $\{ r_{i \to j} \}$ as
\begin{align}
r_{i \to j} = 1 - P_S^{i}(0) \prod_{k \in \partial^a_i \backslash j }  \big[ 1 - p_{ki} \cdot r_{k \to i} \big]. \label{eq:r_i_to_j_cav}
\end{align}
Similarly, we obtain the marginal probability that node $i$ has been infected as
\begin{align}
r_{i} & = 1 - P_S^{i}(\infty) \nonumber \\
& = 1 - P_S^{i}(0) \prod_{k \in \partial^a_i }  \big[ 1 - p_{ki} \cdot r_{k \to i} \big]. \label{eq:r_i_prob}
\end{align}
%
To draw the relation to bond percolation, we consider the scenario where the epidemic outbreaks originate from a single initially infected seed $i_0$ (i.e., the patient zero), such that $P_S^{i_0}(0) = 0, P_I^{i_0}(0) = 1$. Then the set of all nodes that have been infected constitute a giant connected cluster, as every node in this set must have an infection path that starts from $i_0$.
One can interpret $r_{k \to i}$ in Eq.~(\ref{eq:r_i_to_j_cav}) as the cavity probability where node $k$ is in the giant cluster in the absence of node $i$. Upon obtaining the self-consistent solution of Eq.~(\ref{eq:r_i_to_j_cav}) for $\{ r_{k \to i} \}$, one can compute the marginal probability that each node $i$ belongs to the giant cluster according to Eq.~(\ref{eq:r_i_prob}). The size of the giant cluster can then be computed as $\frac{1}{N}\sum_i r_i$, which is the central quantity of interest in percolation problems.

\section*{Supplementary Note 4: Resource Capacity Constraint in Mirror Descent}\label{sec:app_capacity_constr}
When designing algorithms for mitigating network failure, we impose the realistic resource capacity constraint in the form of 
\begin{align}
\sum_{i \in V_b} \sum_{t = 0}^{T-1} \gamma_i(t) \leq \gamma^{\text{tot}}. \label{eq:constr_sum_gamma_in_app}
\end{align}

Such a linear equality constraint can generally be handled by introducing a Lagrangian multiplier and deriving the corresponding Karush-Kuhn-Tucker condition (KKT condition), or by introducing a barrier function as in the interior point method. Consider the latter approach by augmenting the objective function with a log barrier function as
\begin{align}
f(\gamma) & := \gamma^{\text{tot}} - \sum_{i \in V_b} \sum_{t = 0}^{T-1} \gamma_i(t), \\
\mathcal{O}_{\text{aug}} (\gamma) & := \mathcal{O} (\gamma) - \lambda \cdot \log \big( f(\gamma)  \big),
\end{align}
where $\lambda > 0$ is a tunable parameter. The log barrier function strongly penalizes $\mathcal{O}_{\text{aug}} (\gamma)$ when $f(\gamma)$ is close to zero, which encourages $\gamma$ to stay in the interior of the feasible region of Eq.~(\ref{eq:constr_sum_gamma_in_app}).

The gradient of the augmented objective function reads
\begin{align}
\nabla_\gamma \mathcal{O}_{\text{aug}} (\gamma) = \nabla_\gamma \mathcal{O}(\gamma) + \frac{\lambda}{f(\gamma)} \boldsymbol{1}, \label{eq:D_O_aug}
\end{align}
where $\boldsymbol{1}$ is the all-one vector. Eq.~(\ref{eq:D_O_aug}) suggests a global shift of the gradient to encourage the satisfaction of the capacity constraint. This is the motivation for considering a shifted gradient $g^n = \nabla_\gamma \mathcal{O}(\gamma^n) - b^n$ in the mirror descent algorithm in the main text. The gradient shift $-b^n$ is toggled on when $\sum_{t,i}\gamma_{i}^{n} (t) \lessapprox \gamma^{\text{tot}}$, where the shift magnitude $b^n$ is chosen based on the following arguments
\begin{align}
\gamma^{n+1} & =\Psi^{-1}\big(\Psi(\gamma^{n}) - s\cdot\big[\nabla_{\gamma} \mathcal{O} (\gamma^{n})-b^{n}\big]\big), \label{eq:MD_gamma_shifted_gradient} \\
\sum_{t,i}\gamma_{i}^{n+1}(t) & \approx \sum_{t,i}\Psi^{-1}\bigg(\Psi(\gamma_{i}^{n}(t))\bigg) - s \sum_{t,i}\Psi^{-1\prime}\bigg(\Psi(\gamma_{i}^{n}(t))\bigg) \bigg[\frac{\partial}{\partial\gamma_{i}(t)} \mathcal{O} (\gamma^{n})-b^{n}\bigg] \nonumber \\ 
& =\sum_{t,i}\gamma_{i}^{n}(t) - s\sum_{t,i}\gamma_{i}^{n}(t)(1-\gamma_{i}^{n}(t)) \bigg[\frac{\partial}{\partial\gamma_{i}(t)} \mathcal{O} (\gamma^{n})-b^{n}\bigg] \lessapprox \gamma^{\text{tot}},
\end{align}
where a small step size $s$ is assumed.
It requires that
\begin{align}
& \sum_{t,i}\gamma_{i}^{n}(t)(1-\gamma_{i}^{n}(t))\bigg[\frac{\partial}{\partial\gamma_{i}(t)} \mathcal{O} (\gamma^{n})-b^{n}\bigg] \approx 0, \\
& \Longleftrightarrow \quad b^n \approx \frac{\sum_{t,i}\gamma_i^n(t)(1-\gamma_i^n(t))\frac{\partial}{\partial\gamma_i(t)} \mathcal{O} (\gamma^{n})}{\sum_{t,i}\gamma_i^n(t)(1-\gamma_i^n(t))},
\end{align}
which explains the choice of $b^n$ in the main text.

\section*{Supplementary Note 5: Online Supply of Resources}~\label{sec:online_resource}
In this appendix, we consider the cases where the resources arrive in an online fashion, e.g., a limited number of vaccines can be produced every day. In these cases, there is a resource capacity constraint $\gamma_t^{\text{tot}}$ at each time step $t$, yielding the following constrained optimization problems
\begin{align}
& \min_{ \gamma } \;\; \mathcal{O}(\gamma)  = \frac{1}{N} \sum_{i \in V_b} P_F^i(T),  \\
& \text{s. t. } \;\; 0 \leq \gamma_i(t) \leq 1 \quad \forall i, t,  \\
& \qquad \;\; \sum_{i \in V_b} \gamma_i(t) \leq \gamma_t^{\text{tot}} \quad \forall t. \label{eq:constr_sum_gamma_each_t}
\end{align}
The mirror descent algorithm introduced in the main text can be readily applied to solve this optimization problem, except that the shifted gradient $g^n_t = \nabla_{\gamma(t)} \mathcal{O}(\gamma^n(t)) - b^n_t$ has a time-dependent shift $b^n_t$ in this case. The shift magnitude $b^n_t$ at time $t$ can be derived in the same spirit as in Supplementary Note 4, which admits the following expression
\begin{align}
b^n_t \approx \frac{\sum_{i}\gamma_i^n(t)(1-\gamma_i^n(t))\frac{\partial}{\partial\gamma_i(t)} \mathcal{O} (\gamma^{n})}{\sum_{i}\gamma_i^n(t)(1-\gamma_i^n(t))}.
\end{align}
If the resource capacity constraint for $\gamma_t^{\text{tot}}$ is still violated, we project the control variables at time $t$ to the feasible region through a simple rescaling
\begin{align}
\gamma^n(t) \leftarrow \frac{\gamma_t^{\text{tot}}}{\sum_{i} \gamma^n_i(t)} \gamma^n(t).
\end{align}

\begin{figure}
\includegraphics[scale=0.4]{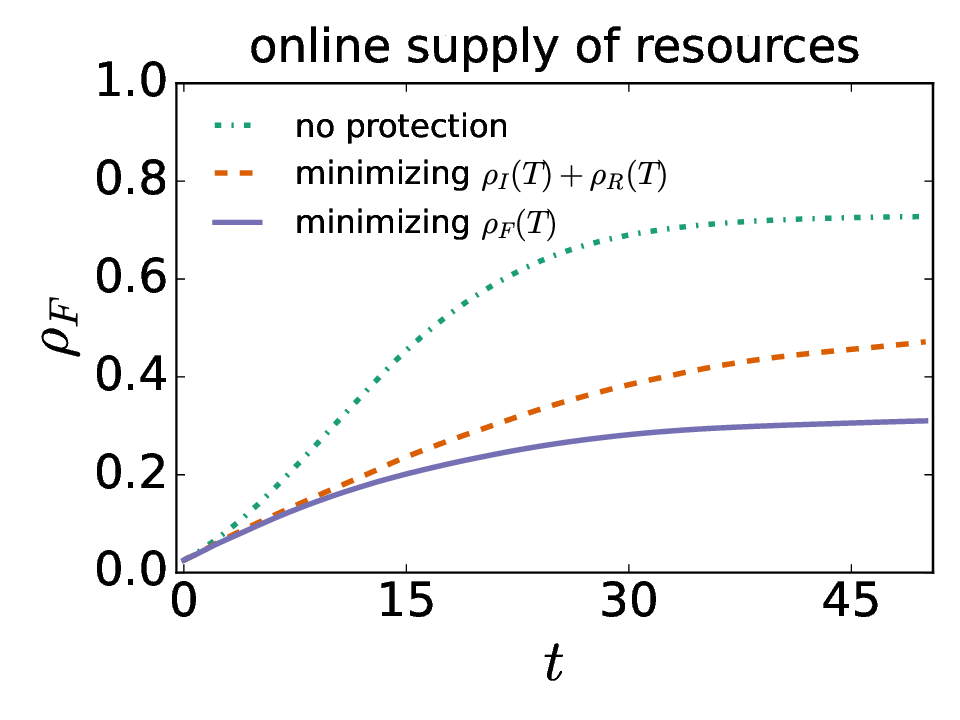}
\caption{Evolution of the size of failures $\rho_F(t)$ under various mitigation strategies. A synthetic two-layer network (mimicing communication network and power grid) is considered. The time window is set as $T = 50$.
The system parameters are $\beta_{ji} = 0.17, \mu = 0.5, \Theta_i = 0.6|\partial_i^b|$. Planted influence parameters $\{ b_{ji} \}$ are considered. Three nodes are randomly chosen as the initially infected individuals, and $\gamma_t^{\text{tot}} = 2$ is considered.}
\label{fig:control_case6}
\end{figure}

As a concrete example, we consider the synthetic two-layer network and the planted influence parameters $\{ b_{ji} \}$ in the subsection ``Case Study in a Synthetic Network'' in Results of the main text. The results of online resource supply in this case are shown in Fig.~\ref{fig:control_case6}, which illustrates the effectiveness of the optimization algorithms for the scenario with the online supply of resources.

\section*{Supplementary Note 6: Probabilistic Seeding}\label{sec:prob_seeding}
There are some cases where the initial infection status of a node $i$ is not fully determined but follows a probability distribution $P_I^i(0)$, which may be obtained after some inference. In the DMP framework, we simply use the available $\{ P_I^i(0) \}$ as the initial condition to iterate the DMP equations, and further optimize the evolution by deploying the protection resources.

\begin{figure}
\includegraphics[scale=1.8]{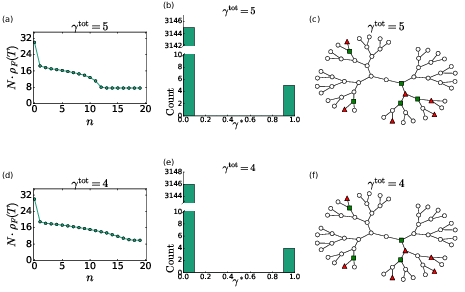}
\caption{Mitigation of the network failures in a binary tree network of size $N=63$ for both layers. Panels (a)(b)(c) correspond to the case with $\gamma^{\text{tot}} = 5$, while Panels (d)(e)(f) correspond to the case with $\gamma^{\text{tot}} = 4$. 
Panels (a) and (d) depict how the final failure size changes during the optimization process. Panels (b) and (e) plot the histogram of the optimal decision variables $\{ \gamma^*_i(t) \}$. Panels (c) and (f) show the optimal placement of resources, where green square nodes receive protection (having a high $\gamma^*_i(t)$ at time $t=0$). The six red triangle nodes are the potential seeds for infection, each of which has an initial infection probability $P_I^i(0) = \frac{1}{2}$. }\label{fig:case_2_control_ab}
\end{figure}

We consider a similar setting as in the subsection ``Case Study in a Tree Network'' of Results in the main text, where there are 3 seeds, each of which has $P_I^i(0) = 1$. In this section, we consider 6 seeds, each of which has $P_I^i(0) = \frac{1}{2}$. The results of optimization are shown in Fig.~\ref{fig:case_2_control_ab}. 
It can be observed that the optimization algorithm can still successfully reduce the final failure size, as in the cases with deterministic seeding.
Interestingly, the optimal protection resource distribution $\{ \gamma^*_i(t) \}$ is also concentrated among a few nodes at a certain time step (as shown in Fig.~\ref{fig:case_2_control_ab}(b) and (e)), even though one cannot be sure about which nodes are initially infected. This may be due to the simple network topology, where there  exists a clear optimal deployment strategy to block the infections coming from the 6 probabilistic seeds altogether, as shown in Fig.~\ref{fig:case_2_control_ab}(c) and (f).

\section*{Supplementary References}
\bibliography{reference}